\documentclass[preprint,showpacs,preprintnumbers,amsmath,amssymb,superscriptaddress]{revtex4}
\usepackage{graphicx}
\usepackage{dcolumn}
\usepackage{amssymb}
\usepackage{amsmath}
\usepackage{bm}
\usepackage[ulem=normalem]{changes}
\usepackage{hyperref}
\usepackage{xfrac}

\begin{document}
\title{Boson Models with Interactions of Arbitrary Order}

\author{P.~Van~Isacker}
\affiliation{Grand Acc\'el\'erateur National d'Ions Lourds,
CEA/DRF - CNRS/IN2P3, Bvd Henri Becquerel, F-14076 Caen, France}

\date{\today}

\begin{abstract}
The paper considers quantal many-boson systems
that are described by a rotationally invariant and boson-number conserving Hamiltonian.
The properties of a generic model are studied which treats
$N$ bosons of $p$ different kinds
with non-zero angular momenta $\ell_1,\ell_2,\dots,\ell_p$,
possibly augmented with a (number of) scalar $s$ boson(s).
The order $k$ of the interaction between the bosons is arbitrary
and closed formulas are given for matrix elements between $N$-boson states
for any $k$ if $p=1$ and $p=2$.
A recursive procedure is defined for arbitrary $k$ and $p$.
With the expressions derived in the paper it is possible to express symbolically
a Hamiltonian matrix element between $N$-boson states
as a linear combination of $k$-body interaction matrix elements.
More generally, the formulas allow the evaluation of matrix elements
of tensor operators that are not necessarily scalar nor boson-number conserving.
The numerical implementation of the formalism is discussed
and illustrated with a few examples.
\end{abstract}

\maketitle

\section{Introduction}
\label{s_intro}
Many quantal systems can be modelled in terms of bosons.
Although the system's basic constituents are fermions
(e.g., neutrons and protons in an atomic nucleus)
clusters of an even number of them can be treated approximately as bosons,
leading to a significant simplification of the original many-fermion problem.
However, since the bosons are not elementary
and violate the Pauli principle that acts among the fermions,
the price to pay is an interaction between the bosons
that is complicated and usually of higher order.

The purpose of this paper is to present a formalism
that deals with quantal systems consisting of bosons
with complex higher-order interactions.
It is assumed that the system is described by a Hamiltonian
that is rotationally invariant
so that its eigenstates are characterised by an angular momentum $J$
and its projection on the $z$ axis $M$.
Furthermore, the Hamiltonian is also assumed to conserve the total number of bosons $N$.
The bosons can be of $p$ different kinds
that have the respective angular momenta $\ell_i, i=1,\dots,p$.
The number $n_i$ of $\ell_i$ bosons is {\em not} conserved in general
and interactions may transform a boson with angular momentum $\ell_i$
into another one with $\ell_{i'}$.
With these assumptions the expressions given in this paper
enable the construction of matrix elements
of a Hamiltonian with $k$-body interactions, or any higher-order non-scalar operator,
between $N$-boson states for arbitrary order $k$
and arbitrary number $p$ of different kinds of bosons.

A system of $N$ bosons of one kind with angular momentum $\ell$
can be considered as the most `elementary' one.
Models of this type shall be referred to as $\ell$-interacting boson model or $\ell$-IBM.
In Sect.~\ref{s_sl} a generic algorithm is presented,
which enables to solve the eigenvalue problem in $\ell$-IBM
for interactions between the bosons that are, in principle, of arbitrary order $k$.

The elementary $\ell$-IBM can be generalised to $\ell_1\dots\ell_p$-IBM.
For $p=2$ explicit formulas are given in Sect.~\ref{s_sll}
for the $N$-boson matrix elements in $\ell_1\ell_2$-IBM,
valid for arbitrary angular momenta $\ell_1$ and $\ell_2$
and arbitrary order $k$ of the interaction.
For $p>2$ explicit formulas become cumbersome
because of the complicated angular momentum recoupling.
In that case a recursive procedure is defined in Sect.~\ref{s_slp},
reducing a matrix element in $\ell_1\dots\ell_p$-IBM
to a linear combination of products of $p$ matrix elements in $\ell_i$-IBM, $i=1,\dots,p$.

The set of bosons $\{b_{\ell_i},i=1,\dots,p\}$
may contain a scalar $s$ boson with zero intrinsic angular momentum, $\ell=0$. 
This possibility covers two important models,
namely the vibron model of molecules~\cite{Iachello95} with $s$ and $p$ ($\ell=1$) bosons
and the IBM of nuclei~\cite{Iachello87} with $s$ and $d$ ($\ell=2$) bosons.
As will be shown below, the $s$-boson dependence of matrix elements can be factored out
if operators are written in normal-ordered form
with creation operators to the left and annihilation operators to the right.
In case the model contains several kinds of $s$ bosons,
they can all be eliminated from normal-ordered operators,
leaving a model of the type $\ell_1\dots\ell_p$-IBM with $\ell_i\neq0$.
The assumption $\ell_i\neq0$ is not a fundamental restriction but is made for convenience.
All formulas given in this paper are valid if some $\ell_i$ are zero;
the recursive algorithm, however, is more efficient
if the $s$ boson(s) is (are) eliminated from the start.

The straightforward elimination of scalar bosons is not possible
if operators are not given in normal-ordered form.
A particularly important case concerns operators given in multipole form,
which is discussed in Sect.~\ref{s_mul}.
An operator in multipole form
can always be reduced to a combination of normal-ordered operators.
In Sect.~\ref{s_mulnor} a generic algorithm is described,
which converts a multipole Hamiltonian into its normal-ordered representation
for arbitrary order $k$ of the interaction.

The formalism presented in this paper is capable of dealing with systems consisting of bosons
that are intrinsically different but have the same angular momentum.
For example, the neutron--proton IBM or IBM-2~\cite{Arima77}
can be considered as $s_\nu s_\pi d_\nu d_\pi$-IBM,
which after the elimination of the $s_\nu$ and $s_\pi$ bosons reduces to $d_\nu d_\pi$-IBM,
the matrix elements of which are quadratic combinations of those in $d$-IBM.
Likewise, the isospin-invariant IBM or IBM-3~\cite{Elliott80}
can be treated as $s_\nu s_\delta s_\pi d_\nu d_\delta d_\pi$-IBM
and can be reduced to a problem in $d$-IBM.

The paper starts with a brief review, given in the next section, of the algorithm
to calculate coefficients of fractional parentage (CFPs) in a seniority basis.

\section{Coefficients of fractional parentage in a seniority basis}
\label{s_cfp}
The definition of CFPs for fermions can be found in standard textbooks~\cite{Shalit63}
and is easily extended to bosons~\cite{Talmi93}.
All formulas given in this section are valid for fermions as well as for bosons
and the notation $j$ is used for the angular momentum of either a fermion or a boson.
Therefore, $j$ can be integer or half-odd-integer.
Sections after the present one only deal with bosons with integer angular momenta $\ell_i$.

A many-body wave function expanded in terms of CFPs
is rotationally invariant, that is, it carries the quantum number of angular momentum $J$,
and it satisfies the requirement of anti-symmetry for fermions or symmetry for bosons.
The definition of CFPs usually applies to single-$j$ systems,
when all particles have the same angular momentum $j$.
This is not a fundamental restriction
as the notion of CFPs can be generalised to multi-$j$ systems~\cite{Skouras86},
consisting of particles with several angular momenta $j_i,i=1,\dots,p$.

A many-body calculation in terms of CFPs has the marked advantage
that the order of the interaction between the particles is automatically taken care of.
This will be illustrated below for single-$j$ CFPs
but the statement is also valid for the multi-$j$ generalisation of CFPs
appropriate for several angular momenta $j_i$.
It is indeed possible to formulate the many-body problem
to any order of the interaction between the particles in terms of generalised CFPs.
This technique can be applied to the shell model in a space with several single-particle levels
or to IBMs with bosons with different angular momenta.
Owing to the proliferation of the number of multi-$j$ CFPs,
this method, however, is restricted to a relatively small number $N$ of particles
and a limited number $p$ of angular momenta $j_i$.

Numerical calculations are vastly more efficient for single-$j$ than they are for multi-$j$ CFPs.
In fact, the matrix elements of a many-body Hamiltonian
can always be written in terms of single-$j$ CFPs
but this requires the recoupling of the angular momenta $j_i$ of the particles.
The analytic expressions of such matrix elements become increasingly cumbersome
as the order $k$ of the interaction as well as the number $p$ of angular momenta $j_i$ increases.
One concludes therefore that many-body calculations in terms of multi-$j$ CFPs
are convenient if the Hamiltonian contains interactions of higher order $k$
but become computationally prohibitively expensive
as the number $N$ of particles and/or the number $p$ of angular momenta increases.
On the other hand, 
analytic expressions for Hamiltonian matrix elements in terms of single-$j$ CFPs
become prohibitively complicated with increasing $k$ or $p$.

The method proposed in this paper is based on single-$j$ CFPs,
which is computationally more efficient than that based on multi-$j$ CFPs.
It avoids the complex analytic expressions of angular-momentum recoupling
through a recursive procedure.

The computation of $(n-1)\times1\rightarrow n$ CFPs
relies on the following recursive formula,
known as the Redmond relation~\cite{Shalit63,Talmi93}:
\begin{align}
[j^{n-1}(\alpha_1J_1),j|\}j^n[\alpha_2J_2]J]={}&
\biggl[
\delta_{\alpha_1\alpha_2}\delta_{J_1J_2}+
(n-1)(-)^{J_1+J_2}[J_1][J_2]
\sum_{\alpha'_1J'_1}
\biggl\{\begin{array}{ccc}J'_1&j&J_1\\J&j&J_2\end{array}\biggr\}
\nonumber\\&\times
[j^{n-2}(\alpha'_1J'_1),j|\}j^{n-1}\alpha_1J_1]
[j^{n-2}(\alpha'_1J'_1),j|\}j^{n-1}\alpha_2J_2]
\biggr],
\label{e_redmond}
\end{align}
where $[x]$ stands for $\sqrt{2x+1}$
and the symbol in curly brackets is a $6j$ symbol.
The recurrence relation~(\ref{e_redmond}) expresses an $(n-1)\times1\rightarrow n$ CFP
on the left-hand side of the equation
in terms of a sum of $(n-2)\times1\rightarrow(n-1)$ CFPs on the right-hand side.
The algorithm cannot be applied as such, however,
since in the CFP on the left-hand side
the $n$-particle state refers to a non-orthonormal, overcomplete basis,
characterised by the labels $[\alpha_2J_2]$ of the $(n-1)$-particle parent state,
while in the CFPs on the right-hand side
the $(n-1)$-particle states must form an orthonormal basis.
To carry out the orthonormalisation,
one orders states in the overcomplete, non-orthonormal basis in a sequence
$|j^n[\alpha_kJ_k]J\rangle$, $k=1,2,\dots,q$,
where $q$ is the number of independent \mbox{(anti-)symmetric} $n$-particle states
with total angular momentum $J$.
The sequence and the order of the states is arbitrary
but for the fact that the states must be linearly independent.
A set of orthonormal states can be obtained from the expansion
\begin{equation}
|j^n\alpha_rJ\rangle=
\sum_{k=1}^ra_{rk}
|j^n[\alpha_kJ_k]J\rangle,
\quad
r=1,2,\dots,q,
\label{e_gs}
\end{equation}
where the coefficients $a_{rk}$ are identified
with the following sequence in a Gram--Schmidt orthonormalisation procedure:
\begin{align}
|j^n\alpha_1J\rangle={}&
\frac{1}{\sqrt{o_{11}}}|j^n[\alpha_1J_1]J\rangle,
\nonumber\\
|j^n\alpha_2J\rangle={}&
\frac{1}{\sqrt{o_{22}-(\tilde o_{21})^2}}
\bigl(|j^n[\alpha_2J_2]J\rangle-
\tilde o_{21}|j^n[\alpha_1J_1]J\rangle\bigr),
\nonumber\\ \vdots{}&\nonumber\\
|j^n\alpha_kJ\rangle={}&
\frac{1}{\sqrt{{\cal N}_k}}
\biggl(|j^n[\alpha_kJ_k]J\rangle-
\sum_{l=1}^{k-1}\tilde o_{kl}|j^n[\alpha_lJ_l]J\rangle\biggr),
\label{e_gsseq}
\end{align}
until $k=q$, with
\begin{equation}
{\cal N}_k=o_{kk}-\sum_{r=1}^{k-1}(\tilde o_{kr})^2,
\quad
o_{kl}=\langle j^n[\alpha_kJ_k]J|j^n[\alpha_lJ_l]J\rangle,
\quad
\tilde o_{kr}=\langle j^n[\alpha_kJ_k]J|j^n\alpha_rJ\rangle.
\label{e_gsnorm}
\end{equation}
The coefficients $a_{rk}$ (needed for $r\geq k$)
can be calculated recursively from
\begin{equation}
a_{kk}=\frac{1}{\sqrt{{\cal N}_k}},
\quad
a_{rk}=
-\frac{1}{\sqrt{{\cal N}_r}}
\sum_{s=k}^{r-1}\tilde o_{rs}a_{sk},
\quad
\tilde o_{rs}=
\sum_{s'=1}^sa_{ss'}o_{rs'},
\label{e_gsa}
\end{equation}
which as only input requires
the knowledge of the overlap matrix
\begin{equation}
o_{kl}=
\sum_{\alpha_iJ_i}
[j^{n-1}(\alpha_iJ_i),j|\}j^n[\alpha_kJ_k]J]
[j^{n-1}(\alpha_iJ_i),j|\}j^n[\alpha_lJ_l]J].
\label{e_over1}
\end{equation}
With the help of Eq.~(\ref{e_gs}),
the CFPs in the orthonormal basis can be expressed
in terms of those in the non-orthonormal basis,
\begin{equation}
[j^{n-1}(\alpha_1J_1),j|\}j^n\alpha_rJ]=
\sum_{k=1}^ra_{rk}
[j^{n-1}(\alpha_1J_1),j|\}j^n[\alpha_kJ_k]J],
\quad
r=1,2,\dots,q.
\label{e_cfp1ort}
\end{equation}

The Redmond relation~(\ref{e_redmond}), together with Eq.~(\ref{e_cfp1ort}),
defines a genuine recursive scheme.
The basis defined by the label $\alpha_r$, however,
depends on the chosen sequence $|j^n[\alpha_kJ_k]J\rangle$, $k=1,2,\dots,q$,
which in general is different from a seniority basis.
The latter basis is mathematically convenient
since it can be obtained from group-theoretical considerations,
as will be discussed in Sect.~\ref{s_sl} in the case of bosons.
In addition, the seniority label has a useful intuitive interpretation
since it corresponds to the number of particles
not in pairs coupled to angular momentum $J=0$~\cite{Racah43}.
For these reasons it is of interest to adapt the Redmond recursive scheme
to include seniority.

To construct basis states $|j^n\upsilon\alpha J\rangle$,
where $\alpha$ is now a label additional to the seniority $\upsilon$,
one proceeds as follows.
As the procedure is recursive,
one assumes that the CFPs $[j^{k-1}(\upsilon_1\alpha_1J_1)j|\}j^k\upsilon\alpha J]$
are known for $k<n$.
The task is to compute them for $k=n$.
All CFPs $[j^{n-1}(\upsilon_1\alpha_1J_1)j|\}j^n\upsilon\alpha J]$ with $\upsilon=n-2,n-4,...,1$ or 0
can be obtained from seniority reduction formulas,
which for fermions ($j$ half-odd-integer) read~\cite{Shalit63}
\begin{align}
&[j^{n-1}(\upsilon-1,\alpha_1J_1),j|\}j^n\upsilon\alpha J]
\nonumber\\&\qquad=
\biggl[\frac{\upsilon(2j+3-n-\upsilon)}{n(2j+3-2\upsilon)}\biggr]^{1/2}
[j^{\upsilon-1}(\upsilon-1,\alpha_1J_1),j|\}j^\upsilon\upsilon\alpha J],
\nonumber\\
&[j^{n-1}(\upsilon+1,\alpha_1J_1),j|\}j^n\upsilon\alpha J]
\nonumber\\&\qquad=
(-)^{J+j-J_1}
\biggl[\frac{(\upsilon+1)(n-\upsilon)(2J_1+1)}{n(2j+1-2\upsilon)(2J+1)}\biggr]^{1/2}
[j^\upsilon(\upsilon\alpha J),j|\}j^{\upsilon+1}\upsilon+1,\alpha_1J_1],
\label{e_senredf}
\end{align}
while for bosons ($j$ integer) they are~\cite{Talmi93}
\begin{align}
&[j^{n-1}(\upsilon-1,\alpha_1J_1),j|\}j^n\upsilon\alpha J]
\nonumber\\&\qquad=
\biggl[\frac{\upsilon(2j-1+n+\upsilon)}{n(2j-1+2\upsilon)}\biggr]^{1/2}
[j^{\upsilon-1}(\upsilon-1,\alpha_1J_1),j|\}j^\upsilon\upsilon\alpha J],
\nonumber\\
&[j^{n-1}(\upsilon+1,\alpha_1J_1),j|\}j^n\upsilon\alpha J]
\nonumber\\&\qquad=
(-)^{J+J_1}
\biggl[\frac{(\upsilon+1)(n-\upsilon)(2J_1+1)}{n(2j+1+2\upsilon)(2J+1)}\biggr]^{1/2}
[j^\upsilon(\upsilon\alpha J),j|\}j^{\upsilon+1}\upsilon+1,\alpha_1J_1].
\label{e_senredb}
\end{align}
Equations~(\ref{e_senredf}) and (\ref{e_senredb}) determine all $(n-1)\times 1\rightarrow n$ CFPs
except those with $\upsilon=n$,
\begin{equation}
[j^{n-1}(\upsilon_1=n-1,\alpha_1J_1),j|\}j^n\upsilon=n,\alpha J].
\label{e_cfpvn1}
\end{equation}
Since a single particle can change seniority by at most one unit,
the $(n-1)$-particle state must have $\upsilon_1=n-1$
in order to lead to an $n$-particle state with $\upsilon=n$.
The CFP~(\ref{e_cfpvn1}) must be obtained from the Redmond relation~(\ref{e_redmond})
and therefore derived from the CFPs in the non-orthonormal basis,
\begin{equation}
[j^{n-1}(\upsilon_1=n-1,\alpha_1J_1),j|\}j^n[\upsilon_2=n-1,\alpha_2J_2]J].
\label{e_cfpvn2}
\end{equation}
The state $|j^n[n-1,\alpha_2J_2]J\rangle$, 
which has the parent $|j^{n-1}n-1,\alpha_2J_2\rangle$,
is a mixture of $\upsilon=n$ and $\upsilon=n-2$ states,
and the latter components must be projected out in order to arrive at an orthonormal basis.
This can be achieved by including the states $|j^nn-2,\alpha J\rangle$
in the Gram--Schmidt orthonormalisation.
The overlap matrix that should be used in this procedure is
\begin{equation}
\biggl(\begin{array}{cc}
\langle j^nn-2,\alpha_rJ|j^nn-2,\alpha_sJ\rangle&
\langle j^nn-2,\alpha_rJ|j^n[n-1,\alpha_lJ_l]J\rangle\\
\langle j^n[n-1,\alpha_kJ_k]J|j^nn-2,\alpha_sJ\rangle&
\langle j^n[n-1,\alpha_kJ_k]J|j^n[n-1,\alpha_lJ_l]J\rangle
\end{array}\biggr),
\label{e_over2}
\end{equation}
where the upper-left matrix is diagonal,
\begin{equation}
\langle j^nn-2,\alpha_rJ|j^nn-2,\alpha_sJ\rangle=\delta_{rs},
\label{e_over3}
\end{equation}
the off-diagonal matrices are
\begin{align}
&\langle j^nn-2,\alpha_rJ|j^n[n-1,\alpha_kJ_k]J\rangle=
\langle j^n[n-1,\alpha_kJ_k]J|j^nn-2,\alpha_rJ\rangle
\nonumber\\&\qquad=
\sum_{\upsilon_i\alpha_iJ_i}
[j^{n-1}(\upsilon_i\alpha_iJ_i),j|\}j^nn-2,\alpha_rJ]
[j^{n-1}(\upsilon_i\alpha_iJ_i),j|\}j^n[n-1,\alpha_kJ_k]J],
\label{e_over4}
\end{align}
and the lower-right matrix is given by Eq.~(\ref{e_over1}).

The above formulas define a recursive procedure
to determine the $(n-1)\times1\rightarrow n$ CFPs
$[j^{n-1}(\upsilon_1\alpha_1J_1),j|\}j^n\upsilon\alpha J]$
in an orthonormal basis that includes the seniority quantum number.
For the construction of the Hamiltonian matrix of a $k$-body interaction in an $n$-particle basis,
knowledge of $k\times r\rightarrow n$ CFPs, with $k+r=n$, is required.
These can be computed from
\begin{align}
&[j^k(\upsilon_k\alpha_kJ_k),j^r(\upsilon_r\alpha_rJ_r)|\}j^n\upsilon_n\alpha_nJ_n]
\nonumber\\&\quad=
[J_k]\sum_{\upsilon'_n\alpha'_nJ'_n}
[J'_n][j,j^{n-1}(\upsilon'_n\alpha'_nJ'_n)|\}j^n\upsilon_n\alpha_nJ_n]
\sum_{\upsilon'_k\alpha'_kJ'_k}
(-)^{j+J'_k+J_n+J_r}
\biggl\{\begin{array}{ccc}j&J'_k&J_k\\J_r&J_n&J'_n\end{array}\biggr\}
\nonumber\\&\qquad\times
[j,j^{k-1}(\upsilon'_k\alpha'_kJ'_k)|\}j^k\upsilon_k\alpha_kJ_k]\,
[j^{k-1}(\upsilon'_k\alpha'_kJ'_k)j^r(\upsilon_r\alpha_rJ_r)|\}j^{n-1}\upsilon'_n\alpha'_nJ'_n].
\label{e_cfport}
\end{align}
This equation is written in terms of $1\times(n-1)\rightarrow n$ CFPs
and defines a recursive procedure ending in $1\times(n-1)\rightarrow n$ CFPs.
These can be related to $(n-1)\times1\rightarrow n$ CFPs through
\begin{equation}
[j,j^{n-1}(\upsilon'_n\alpha'_nJ'_n)|\}j^n\upsilon_n\alpha_nJ_n]=
\phi(-)^{j+J'_n-J_n}[j^{n-1}(\upsilon'_n\alpha'_nJ'_n),j|\}j^n\upsilon_n\alpha_nJ_n],
\end{equation}
where $\phi=+1$ for bosons and $\phi=(-)^{n-1}$ for fermions.

\section{The $s\ell$-IBM}
\label{s_sl}
This section deals with boson models
that include a scalar $s$ boson
and another type of boson with angular momentum $\ell$.
The ensuing model will be referred to as $s\ell$-IBM
and covers two important cases,
namely the vibron model~\cite{Iachello95} with $s$ and $p$ bosons
and the IBM~\cite{Iachello87} with $s$ and $d$ bosons.
As the dependence of matrix elements on the $s$ boson can be readily factored out,
the problem can be reduced to one in $\ell$-IBM.
Note that there is, in principle, no restriction on the values of $\ell$
and therefore the formalism of this section can also be used
to deal with models of interacting bosons with high angular momentum,
such as those related to aligned neutron--proton pairs~\cite{Zerguine11}.

States with $N$ bosons in $s\ell$-IBM can be constructed in the `vibrational' basis
\begin{equation}
\begin{array}{cccccccccc}
{\rm U}(2\ell+2)&\supset&{\rm U}(2\ell+1)&\supset&{\rm SO}(2\ell+1)&
\supset&{\rm SO}(3)&\supset&{\rm SO}(2)\\
\downarrow&&\downarrow&&\downarrow&&\downarrow&&\downarrow\\[0pt]
[N]&&n&&\upsilon&&\alpha L&&M
\end{array},
\label{e_slbasis}
\end{equation}
where $N$ is the total number of $s$+$\ell$ bosons
and the number of $\ell$ bosons is $n=0,1,\dots,N$.
The states in the $\ell^n$ system associated with the classification 
${\rm U}(2\ell+1)\supset{\rm SO}(2\ell+1)\supset{\rm SO}(3)$
coincide with those obtained with the CFP algorithm explained in the previous section.
The basis states~(\ref{e_slbasis}) are characterised by the seniority quantum number $\upsilon$,
which can take the values $\upsilon=n,n-2,\dots,0$ or 1.
The allowed values of the angular momentum $L$
follow from the \mbox{${\rm SO}(2\ell+1)\supset{\rm SO}(3)$} reduction,
for which no simple, closed formula exists for general $\ell$
but which can be obtained with group-theoretical methods~\cite{Wybourne70}.
The $\ell$ bosons are possibly characterised by an additional label $\alpha$,
which specifies how many times the angular momentum $L$
occurs for a given seniority $\upsilon$, $\alpha=1,2,\dots,d_\ell(\upsilon,L)$.
The multiplicity $d_\ell(\upsilon,L)$ is known
as a complex integral over the characters of SO($2\ell+1$) and SO(3)~\cite{Weyl39,Gheorghe04},
\begin{equation}
d_\ell(\upsilon,L)=
\frac{i}{2\pi}\oint_{|z|=1}
\frac{(z^{2L+1}-1)(z^{2\upsilon+2\ell-1}-1)\prod_{k=1}^{2\ell-2}(z^{\upsilon+k}-1)}
{z^{\ell\upsilon+L+2}\prod_{k=1}^{2\ell-2}(z^{k+1}-1)}dz.
\label{e_slmult}
\end{equation}
For simplicity of notation, the label $\alpha$ will be suppressed in the following
if it is not needed, that is, if $d_\ell(\upsilon,L)=1$.
The label $M$ of the final SO(2) algebra in Eq.~(\ref{e_slbasis})
is the projection of the angular momentum $L$ on the $z$ axis,
which satisfies $M=-L,-L+1,\dots,+L$.
Throughout this paper rotationally invariant Hamiltonians are considered,
for which all such $M$ states have the same energy.
The label $M$ will be suppressed if it is not needed.

A general number-conserving $k$-body interaction
between the bosons can be expressed as follows:
\begin{equation}
\sum_L\sum_{m\nu}\sum_{m'\nu'}
v_{m\nu,m'\nu'}^{kL}
(-)^LB_{km\nu L}^\dag\cdot\tilde B_{km'\nu'L},
\label{e_slkbody}
\end{equation}
where the operator $B_{km\nu LM}^\dag$ ($B_{km\nu LM}$)
creates (annihilates) a normalised state of $k$ bosons according to the classification~(\ref{e_slbasis}),
with $m$ the number of $\ell$ bosons, $\nu$ their seniority
and $L$ their angular momentum.
Further, the dot indicates a scalar product,
$v_{m\nu,m'\nu'}^{kL}$ is a numerical coefficient
that determines the strength of the interaction
and $\tilde B_{km\nu LM}\equiv(-)^{L-M}B_{km\nu L-M}$.
The $s$-boson part in the operators $B_{km\nu LM}^\dag$ and $\tilde B_{km\nu LM}$
can easily be taken care of,
\begin{equation}
B_{km\nu LM}^\dag=
\frac{(s^\dag)^{k-m}}{\sqrt{(k-m)!}}B_{m\nu LM}^\dag,
\quad
\tilde B_{km\nu LM}=
\frac{(\tilde s)^{k-m}}{\sqrt{(k-m)!}}\tilde B_{m\nu LM},
\label{e_slnbody}
\end{equation}
where $B_{m\nu LM}^\dag$ ($\tilde B_{m\nu LM}$) creates (annihilates)
a normalised state of $m$ $\ell$ bosons.

Matrix elements of an interaction term in the expansion~(\ref{e_slkbody})
between basis states~(\ref{e_slbasis}) can be factored as follows:
\begin{align}
&\langle Nn\upsilon\alpha J|
(-)^LB_{km\nu L}^\dag\cdot\tilde B_{km'\nu'L}
|Nn'\upsilon'\alpha'J\rangle
\nonumber\\&\qquad=
\frac{\langle s^{N-n}|(s^\dag)^{k-m}(\tilde s)^{k-m'}|s^{N-n'}\rangle}{\sqrt{(k-m)!(k-m')!}}
\langle n\upsilon\alpha J|
(-)^LB_{m\nu L}^\dag\cdot\tilde B_{m'\nu'L}
|n'\upsilon'\alpha'J\rangle.
\label{e_slmes1}
\end{align}
The $s$-boson matrix element in this equation follows from
\begin{equation}
\frac{\langle s^{n_s}|(s^\dag)^{m_s}(\tilde s)^{m'_s}|s^{n'_s}\rangle}{\sqrt{m_s!m'_s!}}=
\biggl[\biggl(\begin{array}{c}n_s\\m_s\end{array}\biggr)
\biggl(\begin{array}{c}n'_s\\m'_s\end{array}\biggr)\biggr]^{1/2}
\delta_{n_s-m_s,n'_s-m'_s},
\label{e_slmes2}
\end{equation}
while the expression for the $\ell$-boson matrix element is
\begin{align}
&\langle n\upsilon\alpha J|
(-)^LB_{m\nu L}^\dag\cdot\tilde B_{m'\nu'L}
|n'\upsilon'\alpha'J\rangle=
\biggl[\biggl(\begin{array}{c}n\\m\end{array}\biggr)
\biggl(\begin{array}{c}n'\\m'\end{array}\biggr)\biggr]^{1/2}
\sum_{n''\upsilon''\alpha''L''}
\delta_{n'',n-m}\delta_{n'',n'-m'}
\nonumber\\&\quad\times
[\ell^{n''}(\upsilon''\alpha''L'')\ell^m(\nu L)|\}\ell^n\upsilon\alpha J]
[\ell^{n''}(\upsilon''\alpha''L'')\ell^{m'}(\nu'L)|\}\ell^{n'}\upsilon'\alpha'J],
\label{e_slmes3}
\end{align}
in terms of the $n''\times m\rightarrow n$ and $n''\times m'\rightarrow n'$  CFPs,
which are known from Eq.~(\ref{e_cfport}).
Note that the multiplicity label $\alpha$ is included in the states with $n$, $n'$ and $n''$ bosons,
where it is frequently needed,
but not in the configuration that specifies the interaction,
where it is usually not required.
For example, in $sd$-IBM no multiplicity label is needed
up to and including five-body interactions.

Matrix elements of non-scalar operators that do not necessarily conserve boson number
can be derived in a similar fashion.
One finds
\begin{align}
&\langle Nn\upsilon\alpha J\|
\bigl(B_{km\nu L}^\dag\times\tilde B_{k'm'\nu'L'}\bigr)^{(R)}
\|N'n'\upsilon'\alpha'J'\rangle
\nonumber\\&\quad=
\frac{\langle s^{N-n}|(s^\dag)^{k-m}(\tilde s)^{k'-m'}|s^{N'-n'}\rangle}{\sqrt{(k-m)!(k'-m')!}}
\langle n\upsilon\alpha J\|
\bigl(B_{m\nu L}^\dag\times\tilde B_{m'\nu'L'}\bigr)^{(R)}
\|n'\upsilon'\alpha'J'\rangle.
\label{e_slrmes1}
\end{align}
The $s$-boson matrix element is given by Eq.~(\ref{e_slmes2})
while for the $\ell$-boson matrix element one obtains
\begin{align}
&\langle n\upsilon\alpha J\|
\bigl(B_{m\nu L}^\dag\times\tilde B_{m'\nu'L'}\bigr)^{(R)}
\|n'\upsilon'\alpha'J'\rangle
\nonumber\\&=
[J][R][J']
\biggl[\biggl(\begin{array}{c}n\\m\end{array}\biggr)
\biggl(\begin{array}{c}n'\\m'\end{array}\biggr)\biggr]^{1/2}
\sum_{n''\upsilon''\alpha''J''}(-)^{J+R+L'+J''}
\biggl\{\begin{array}{ccc}L&L'&R\\J'&J&J''\end{array}\biggr\}
\delta_{n'',n-m}\delta_{n'',n'-m'}
\nonumber\\&\quad\times
[\ell^{n''}(\upsilon''\alpha''J'')\ell^m(\nu L)|\}\ell^n\upsilon\alpha J]
[\ell^{n''}(\upsilon''\alpha''J'')\ell^{m'}(\nu'L)|\}\ell^{n'}\upsilon'\alpha'J'].
\label{e_slrmes2}
\end{align}
The doubled-barred matrix elements in Eqs.~(\ref{e_slrmes1}) and~(\ref{e_slrmes2})
are reduced in angular momentum by virtue of the Wigner--Eckart theorem,
defined according to the convention of Refs.~\cite{Shalit63,Talmi93}.

Equations~(\ref{e_slmes1}--\ref{e_slrmes2})
are closed expressions for matrix elements between many-boson states in $s\ell$-IBM
of a $k$-body interaction and of a non-scalar operator of arbitrary order.
While there is, in principle, no restriction
on the order $k$ of the interaction or on the number of bosons $N$,
the computational cost for calculating the $k\times r\rightarrow n$ CFPs
increases with $k$ and $N$ since these are obtained recursively.

\section{The $s\ell_1\ell_2$-IBM}
\label{s_sll}
This section deals with a system consisting of $N$ bosons that occupy an $s$ level
and two additional levels with angular momenta $\ell_1$ and $\ell_2$.
The most relevant application of this case is $sdg$-IBM~\cite{Devi92}
with $s$, $d$ and $g$ bosons, that is, if $\ell_1=2$ and $\ell_2=4$.
Throughout this section it is assumed that
labels with index `1' act on the $\ell_1$ bosons
while those with index `2'  act on the $\ell_2$ bosons.

States with $N$ bosons in $s\ell_1\ell_2$-IBM can be defined in the basis
\begin{equation}
\begin{array}{ccccccccc}
{\rm U}(\Omega_{12}+1)&\supset&{\rm U}(\Omega_1)&\otimes&
{\rm U}(\Omega_2)&\supset&\cdots&\supset&{\rm SO}_{12}(3)\\
\downarrow&&\downarrow&&\downarrow&&&&\downarrow\\[0pt]
[N]&&n_1&&n_2&&&&L
\end{array},
\label{e_sllbasis1}
\end{equation}
where $\Omega_i=2\ell_i+1$ and $\Omega_{12}=\Omega_1+\Omega_2$.
The number of $\ell_i$ bosons is $n_i$
and their sum is $N$, that is, $(n_1,n_2)=(N,0),(N-1,1),\dots,(0,N)$.
The intermediate classification in Eq.~(\ref{e_sllbasis1}) for each unitary algebra ${\rm U}(\Omega_i)$ reads
\begin{equation}
\begin{array}{cccccc}
{\rm U}(\Omega_i)&\supset&{\rm SO}(\Omega_i)&\supset&{\rm SO}_i(3)\\
\downarrow&&\downarrow&&\downarrow\\[0pt]
n_i&&\upsilon_i&&\alpha_iL_i
\end{array},
\label{e_sllbasis2}
\end{equation}
with labels as defined in the previous section.
It is convenient to introduce at this point the short-hand notation
$\Gamma_i$ for the set of labels $\{n_i\upsilon_i\alpha_iJ_i\}$ associated with a state
and $\Lambda_i$ for labels $\{m_i\nu_iL_i\}$ associated with an operator.

A general number-conserving $k$-body interaction
between the bosons can be expressed as follows:
\begin{equation}
\sum_{LM}\sum_{\Lambda_1\Lambda_2}\sum_{\Lambda'_1\Lambda'_2}
v_{\Lambda_1\Lambda_2,\Lambda'_1\Lambda'_2}^{kL}
\frac{(s^\dag)^{k-m}(\tilde s)^{k-m'}}{\sqrt{(k-m)!(k-m')!}}
\bigl(B_{\Lambda_1}^\dag\times B_{\Lambda_2}^\dag\bigr)^{(L)}_M
{\bigl(B_{\Lambda'_1}^\dag\times B_{\Lambda'_2}^\dag\bigr)^{(L)}_M}^\dag,
\label{e_sllkbody1}
\end{equation}
with $m\equiv m_1+m_2$ and $m'\equiv m'_1+m'_2$.
With the result
\begin{equation}
{\bigl(B_{\Lambda_1}^\dag\times B_{\Lambda_2}^\dag\bigr)^{(L)}_M}^\dag=
(-)^{L-M}\bigl(\tilde B_{\Lambda_1}\times\tilde B_{\Lambda_2}\bigr)^{(L)}_{-M}\,,
\label{e_sllhermit}
\end{equation}
the interaction can be rewritten as
\begin{equation}
\sum_L\sum_{\Lambda_1\Lambda_2}\sum_{\Lambda'_1\Lambda'_2}
v_{\Lambda_1\Lambda_2,\Lambda'_1\Lambda'_2}^{kL}
\frac{(s^\dag)^{k-m}(\tilde s)^{k-m'}}
{\sqrt{(k-m)!(k-m')!}}
(-)^LB_{\Lambda_1\Lambda_2L}^\dag\cdot\tilde B_{\Lambda'_1\Lambda'_2L}\,,
\label{e_sllkbody2}
\end{equation}
where the $B$ operators are defined as
\begin{equation}
B_{\Lambda_1\Lambda_2LM}^\dag\equiv
\bigl(B_{\Lambda_1}^\dag\times B_{\Lambda_2}^\dag\bigr)^{(L)}_M,
\qquad
\tilde B_{\Lambda_1\Lambda_2LM}\equiv
\bigl(\tilde B_{\Lambda_1}\times\tilde B_{\Lambda_2}\bigr)^{(L)}_M.
\label{e_sllboper1}
\end{equation}

Matrix elements of a single interaction term in the expansion~(\ref{e_sllkbody2}),
that is, of the operator
\begin{equation}
\frac{(s^\dag)^{k-m}(\tilde s)^{k-m'}}{\sqrt{(k-m)!(k-m')!}}
(-)^LB_{\Lambda_1\Lambda_2L}^\dag\cdot\tilde B_{\Lambda'_1\Lambda'_2L},
\label{e_sllkbody3}
\end{equation}
between the basis states~(\ref{e_sllbasis1},\ref{e_sllbasis2})
can be written as the product of two factors.
The first one equals
\begin{equation}
\frac{\langle s^{N-n}|(s^\dag)^{k-m}
(\tilde s)^{k-m'}|s^{N-n'}\rangle}
{\sqrt{(k-m)!(k-m')!}},
\label{e_sllmes1}
\end{equation}
with $n\equiv n_1+n_2$ and $n'\equiv n'_1+n'_2$.
This factor is known from Eq.~(\ref{e_slmes2}) while the second is given by
\begin{align}
&\langle N\Gamma_1\Gamma_2;J|
(-)^LB_{\Lambda_1\Lambda_2L}^\dag\cdot\tilde B_{\Lambda'_1\Lambda'_2L}
|N\Gamma'_1\Gamma'_2;J\rangle
\nonumber\\&\qquad=
(-)^{J'_1+J_2+L'_1+L_2+L+J}(2L+1)
\sum_R
\biggl\{\begin{array}{ccc}L_1&L_2&L\\L'_2&L'_1&R\end{array}\biggr\}
\biggl\{\begin{array}{ccc}J_1&J_2&J\\J'_2&J'_1&R\end{array}\biggr\}
\nonumber\\&\qquad\quad\times
\langle\Gamma_1\|
\bigl(B_{\Lambda_1}^\dag\times\tilde B_{\Lambda'_1}\bigr)^{(R)}
\|\Gamma'_1\rangle
\langle\Gamma_2\|
\bigl(B_{\Lambda_2}^\dag\times\tilde B_{\Lambda'_2}\bigr)^{(R)}
\|\Gamma'_2\rangle.
\label{e_sllmes2}
\end{align}
The reduced matrix elements on the right-hand side of Eq.~(\ref{e_sllmes2})
are known from Eq.~(\ref{e_slrmes2}).

Matrix elements of a non-scalar operator can be dealt with in a similar fashion.
One introduces the $B$ operators
\begin{align}
B_{k\Lambda_1\Lambda_2LM}^\dag\equiv{}&
\frac{(s^\dag)^{k-m}}{\sqrt{(k-m)!}}
\bigl(B_{\Lambda_1}^\dag\times B_{\Lambda_2}^\dag\bigr)^{(L)}_M=
\frac{(s^\dag)^{k-m}}{\sqrt{(k-m)!}}
B_{\Lambda_1\Lambda_2LM}^\dag,
\nonumber\\
\tilde B_{k\Lambda_1\Lambda_2LM}\equiv{}&
\frac{(\tilde s)^{k-m}}{\sqrt{(k-m)!}}
\bigl(\tilde B_{\Lambda_1}\times\tilde B_{\Lambda_2}\bigr)^{(L)}_M=
\frac{(\tilde s)^{k-m}}{\sqrt{(k-m)!}}
\tilde B_{\Lambda_1\Lambda_2LM},
\label{e_sllboper2}
\end{align}
and calculates the matrix element of a general non-scalar operator as
\begin{align}
&\langle N\Gamma_1\Gamma_2;J\|
\bigl(B_{k\Lambda_1\Lambda_2L}^\dag\times\tilde B_{k'\Lambda'_1\Lambda'_2L'}\bigr)^{(R)}
\|N'\Gamma'_1\Gamma'_2;J'\rangle
\nonumber\\&\quad=
\frac{\langle s^{N-n}|(s^\dag)^{k-m}(\tilde s)^{k'-m'}|s^{N'-n'}\rangle}{\sqrt{(k-m)!(k'-m')!}}
\langle\Gamma_1\Gamma_2;J\|
\bigl(B_{\Lambda_1\Lambda_2L}^\dag\times\tilde B_{\Lambda'_1\Lambda'_2L'}\bigr)^{(R)}
\|\Gamma'_1\Gamma'_2;J'\rangle.
\label{e_sllrmes1}
\end{align}
The $s$-boson matrix element is given by Eq.~(\ref{e_sllmes1})
while for the $\ell_1\ell_2$-boson matrix element one obtains
\begin{align}
&\langle\Gamma_1\Gamma_2;J\|
\bigl(B_{\Lambda_1\Lambda_2L}^\dag\times\tilde B_{\Lambda'_1\Lambda'_2L'}\bigr)^{(R)}
\|\Gamma'_1\Gamma'_2;J'\rangle
\nonumber\\&\qquad=
[L][J][R][L'][J']\sum_{R_1R_2}[R_1][R_2]
\left\{\begin{array}{ccc}L_1&L_2&L\\L'_1&L'_2&L'\\R_1&R_2&R\end{array}\right\}
\left\{\begin{array}{ccc}J_1&J_2&J\\J'_1&J'_2&J'\\R_1&R_2&R\end{array}\right\}
\nonumber\\&\qquad\quad\times
\langle\Gamma_1\|
\bigl(B_{\Lambda_1}^\dag\times\tilde B_{\Lambda'_1}\bigr)^{(R_1)}
\|\Gamma'_1\rangle
\langle\Gamma_2\|
\bigl(B_{\Lambda_2}^\dag\times\tilde B_{\Lambda'_2}\bigr)^{(R_2)}
\|\Gamma'_2\rangle,
\label{e_sllrmes2}
\end{align}
where the symbols in curly brackets are $9j$ symbols~\cite{Shalit63,Talmi93}.

Equations~(\ref{e_sllmes1},\ref{e_sllmes2}) and~(\ref{e_sllrmes1},\ref{e_sllrmes2})
are closed expressions for matrix elements between many-boson states in $s\ell_1\ell_2$-IBM
of a $k$-body interaction and of a non-scalar operator of arbitrary order.
As a reminder, in these equations
$\Gamma_i$ stands for the labels $\{n_i\upsilon_i\alpha_iJ_i\}$ of a state
and $\Lambda_i$ for the labels $\{m_i\nu_iL_i\}$ of an operator,
both associated with the boson with angular momentum $\ell_i$.

\section{The $s\ell_1\dots\ell_p$-IBM}
\label{s_slp}
As the number of the different kinds of bosons increases
the angular-momentum recoupling becomes more complicated
and the resulting expressions increasingly cumbersome.
Instead of giving closed expressions in terms of $3nj$ symbols,
it is more appropriate to define a recursive procedure,
reducing a calculation in $s\ell_1\dots\ell_p$-IBM
to one in $s\ell_1\dots\ell_{p-1}$-IBM,
as will be explained in this section.

Basis states in $s\ell_1\dots\ell_p$-IBM can be defined recursively according to
\begin{equation}
\begin{array}{ccccccccc}
{\rm U}(\Omega_{1\dots p}+1)&\supset&{\rm U}(\Omega_{1\dots p-1})
&\otimes&{\rm U}(\Omega_p)&\supset&\cdots&\supset&{\rm SO}_{1\dots p}(3)\\
\downarrow&&\downarrow&&\downarrow&&&&\downarrow\\[0pt]
[N]&&N-n_p&&n_p&&&&L
\end{array},
\label{e_slpbasis}
\end{equation}
where $\Omega_i=2\ell_i+1$
and $\Omega_{1\dots r}=\Omega_1+\cdots+\Omega_r$.
The intermediate algebra ${\rm U}(\Omega_{1\dots p-1})$
can be treated in the same fashion, ending in ${\rm SO}_{1\dots p-1}(3)$,
which is coupled with ${\rm SO}_p(3)$
to the total angular momentum algebra ${\rm SO}_{1\dots p}(3)$.
According to this classification an operator
can be defined as follows:
\begin{equation}
\bigl(\cdots\bigl(\bigl(B_{\Lambda_1}^\dag\times B_{\Lambda_2}^\dag\bigr)^{(I_2)}\times
B_{\Lambda_3}^\dag\bigr)^{(I_3)}\times\cdots\times B_{\Lambda_p}^\dag\bigr)^{(L)}_M,
\label{e_slpope1}
\end{equation}
where $\Lambda_i$ stands for the labels $\{m_i\nu_iL_i\}$
and $I_i$ are intermediate angular momenta,
resulting from the coupling of $I_{i-1}$ and $L_i$,
in the convention that $I_1=L_1$
and that $I_p=L$ is the total angular momentum of the operator.
With the help of Eq.~(\ref{e_sllhermit}) it can be shown by induction
that the Hermitian conjugate of the operator~(\ref{e_slpope1}) is
\begin{align}
&{\bigl(\cdots\bigl(\bigl(B_{\Lambda_1}^\dag\times B_{\Lambda_2}^\dag\bigr)^{(I_2)}\times
B_{\Lambda_3}^\dag\bigr)^{(I_3)}\times\cdots\times B_{\Lambda_p}^\dag\bigr)^{(L)}_M}^\dag
\nonumber\\&\quad=
(-)^{L-M}
\bigl(\cdots\bigl(\bigl(\tilde B_{\Lambda_1}\times\tilde B_{\Lambda_2}\bigr)^{(I_2)}\times
\tilde B_{\Lambda_3}\bigr)^{(I_3)}\times\cdots\times\tilde B_{\Lambda_p}\bigr)^{(L)}_{-M}.
\label{e_slpope2}
\end{align}
Therefore, a general representation of a $k$-body interaction in $s\ell_1\dots\ell_p$-IBM is
\begin{equation}
\sum_L\sum_{\bar\Lambda\bar I}\sum_{\bar\Lambda'\bar I'}
v_{\bar\Lambda\bar I,\bar\Lambda'\bar I'}^{kL}
\frac{(s^\dag)^{k-m}(\tilde s)^{k-m'}}
{\sqrt{(k-m)!(k-m')!}}
(-)^LB_{\{\bar\Lambda\bar I\}_pL}^\dag\cdot\tilde B_{\{\bar\Lambda'\bar I'\}_pL}\,,
\label{e_slpkbody}
\end{equation}
where $k-m$ ($k-m'$) is the number of $s$-boson creation (annihilation) operators,
which implies $m\equiv\sum_im_i$ and $m'\equiv\sum_im'_i$,
where the summation is over all $\ell_i$ bosons, $i=1,\dots,p$.
The $B$ operators are defined as
\begin{align}
B_{\{\bar\Lambda\bar I\}_pLM}^\dag\equiv{}&
\bigl(\cdots\bigl(\bigl(B_{\Lambda_1}^\dag\times B_{\Lambda_2}^\dag\bigr)^{(I_2)}\times
B_{\Lambda_3}^\dag\bigr)^{(I_3)}\times\cdots\times B_{\Lambda_p}^\dag\bigr)^{(L)}_M,
\nonumber\\
\tilde B_{\{\bar\Lambda\bar I\}_pLM}\equiv{}&
\bigl(\cdots\bigl(\bigl(\tilde B_{\Lambda_1}\times\tilde B_{\Lambda_2}\bigr)^{(I_2)}\times
\tilde B_{\Lambda_3}\bigr)^{(I_3)}\times\cdots\times\tilde B_{\Lambda_p}\bigr)^{(L)}_M.
\label{e_slpboper}
\end{align}
The labels $\{\Lambda_1,\dots,\Lambda_p\}$ and $\{I_1,\dots,I_p\}$
are collectively denoted as $\{\bar\Lambda\bar I\}_p$.
As the proposed method is recursive in $p$,
it is necessary to indicate the number of labels in the sets $\bar\Lambda$ and $\bar I$.
In the interaction matrix elements $v_{\bar\Lambda\bar I,\bar\Lambda'\bar I'}^{kL}$
this index can be dropped 
since they are defined in $s\ell_1\dots\ell_p$-IBM for a given, fixed $p$.

It is straightforward to deal with the $s$-boson dependence of any matrix element,
which is contained in the factor
\begin{equation}
\frac{\langle s^{N-n}|(s^\dag)^{k-m}
(\tilde s)^{k'-m'}|s^{N'-n'}\rangle}
{\sqrt{(k-m)!(k'-m')!}},
\label{e_slpmes1}
\end{equation}
with $n\equiv\sum_in_i$ and $n'\equiv\sum_in'_i$.
The expression~(\ref{e_slmes2}) can be used for a general non-scalar operator,
with possibly $N\neq N'$ and $k\neq k'$,
and for a $k$-body interaction, for which $N=N'$ and $k=k'$.

After factoring out the $s$-boson dependence,
the remaining matrix element is defined in $\ell_1\dots\ell_p$-IBM.
A basis state of $\ell_1\dots\ell_p$-IBM can be written as
\begin{equation}
|(\cdots((\Gamma_1\times\Gamma_2)^{(K_2)}\times\Gamma_3)^{(K_3)}
\times\cdots\times\Gamma_p)^{(J)}_M\rangle\equiv
|\{\bar\Gamma\bar K\}_p;JM\rangle,
\label{e_slpbas1}
\end{equation}
where $\Gamma_i$ stands for the labels $\{n_i\upsilon_i\alpha_iJ_i\}$.
Akin to the notation for the operators,
the labels $\{\Gamma_1,\dots,\Gamma_p\}$ and $\{K_1,\dots,K_p\}$
are collectively denoted as $\{\bar\Gamma\bar K\}_p$.
The intermediate angular momenta $K_i$
result from the coupling of $K_{i-1}$ with $J_i$,
in the convention that $K_1=J_1$
and that $K_p=J$ is the total angular momentum of the state.
The task is therefore the compute
\begin{equation}
\langle\{\bar\Gamma\bar K\}_p;J=K_p|
(-)^LB_{\{\bar\Lambda\bar I\}_pL}^\dag\cdot\tilde B_{\{\bar\Lambda'\bar I'\}_pL}
|\{\bar\Gamma'\bar K'\}_p;J=K'_p\rangle,
\label{e_slpmes2}
\end{equation}
which is a generic matrix element of an interaction term in the expansion~(\ref{e_slpkbody})
between basis states~(\ref{e_slpbas1}) of $\ell_1\dots\ell_p$-IBM.
With use of the operator representation~(\ref{e_slpope1})
this matrix element can be rewritten as
\begin{align}
&\langle\{\bar\Gamma\bar K\}_p|
(-)^LB_{\{\bar\Lambda\bar I\}_p}^\dag\cdot\tilde B_{\{\bar\Lambda'\bar I'\}_p}
|\{\bar\Gamma'\bar K'\}_p\rangle
\nonumber\\&\quad=
(-)^{K'_{p-1}+J_p+I'_{p-1}+L_p+L+J}(2L+1)
\sum_R
\biggl\{\begin{array}{ccc}I_{p-1}&L_p&I_p\\L'_p&I'_{p-1}&R\end{array}\biggr\}
\biggl\{\begin{array}{ccc}K_{p-1}&J_p&K_p\\J'_p&K'_{p-1}&R\end{array}\biggr\}
\nonumber\\&\qquad\times
\langle\{\bar\Gamma\bar K\}_{p-1}\|
\bigl(B_{\{\bar\Lambda\bar I\}_{p-1}}^\dag\times\tilde B_{\{\bar\Lambda'\bar I'\}_{p-1}}\bigr)^{(R)}
\|\{\bar\Gamma'\bar K'\}_{p-1}\rangle
\langle\Gamma_p\|
\bigl(B_{\Lambda_p}^\dag\times\tilde B_{\Lambda'_p}\bigr)^{(R)}
\|\Gamma'_p\rangle.
\label{e_slpmes2}
\end{align}
The interaction matrix element is expressed
as a sum of products of two reduced matrix elements,
the last one of which is
\begin{equation}
\langle\Gamma_p\|
\bigl(B_{\Lambda_p}^\dag\times\tilde B_{\Lambda'_p}\bigr)^{(R)}
\|\Gamma'_p\rangle=
\langle n_p\upsilon_p\alpha_pJ_p\|
\bigl(B_{m_p\nu_pL_p}^\dag\times\tilde B_{m'_p\nu'_pL'_p}\bigr)^{(R)}
\|n'_p\upsilon'_p\alpha'_pJ'_p\rangle,
\label{e_slprmes1}
\end{equation}
where, for clarity's sake, on the right-hand side all labels
contained in $\Gamma_p$, $\Gamma'_p$, $\Lambda_p$ and $\Lambda'_p$
are denoted explicitly.
The reduced matrix element~(\ref{e_slprmes1}) is known from Eq.~(\ref{e_slrmes2})
while the first reduced matrix element in Eq.~(\ref{e_slpmes2})
is defined in $\ell_1\dots\ell_{p-1}$-IBM.
Note however that, although one started out
with the matrix element of a scalar operator,
the expansion involves reduced matrix elements of non-scalar operators.
The recursive procedure can be readily extended to the latter operators
through the relation
\begin{align}
&\langle\{\bar\Gamma\bar K\}_p\|
\bigl(B_{\{\bar\Lambda\bar I\}_p}^\dag\times\tilde B_{\{\bar\Lambda'\bar I'\}_p}\bigr)^{(R)}
\|\{\bar\Gamma'\bar K'\}_p\rangle
\label{e_slprmes2}\\&\quad=
[I_p][K_p][R][I'_p][K'_p]\sum_{R_{p-1}R_p}[R_{p-1}][R_p]
\left\{\begin{array}{ccc}I_{p-1}&L_p&I_p\\I'_{p-1}&L'_p&I'_p\\R_{p-1}&R_p&R\end{array}\right\}
\left\{\begin{array}{ccc}K_{p-1}&J_p&K_p\\K'_{p-1}&J'_p&K'_p\\R_{p-1}&R_p&R\end{array}\right\}
\nonumber\\&\qquad\times
\langle\{\bar\Gamma\bar K\}_{p-1}\|
\bigl(B_{\{\bar\Lambda\bar I\}_{p-1}}^\dag\times\tilde B_{\{\bar\Lambda'\bar I'\}_{p-1}}\bigr)^{(R_{p-1})}
\|\{\bar\Gamma'\bar K'\}_{p-1}\rangle
\langle\Gamma_p\|
\bigl(B_{\Lambda_p}^\dag\times\tilde B_{\Lambda'_p}\bigr)^{(R_p)}
\|\Gamma'_p\rangle.
\nonumber
\end{align}
This result is the generalisation of Eq.~(\ref{e_sllrmes2}) from $p=2$ to arbitrary $p$.
The total angular momenta $J$, $J'$, $L$ and $L'$
of the states in bra and ket and of the $B^\dag$ and $\tilde B$ operators
coincide with the final angular momenta
in the series $\bar\Gamma$, $\bar K$, $\bar\Lambda$ and $\bar I$,  respectively,
which implies the equivalences $J=K_p$, $J'=K'_p$, $L=I_p$ and $L'=I'_p$.

\section{Multipole interaction to any order}
\label{s_mul}
The previous sections described the evaluation of matrix elements in $\ell_1\dots\ell_p$-IBM
if the Hamiltonian and other operators are specified in normal-ordered form,
that is, with all creation operators on the left
and all annihilation operators on the right.
Alternatively, operators can be written in multipole form in terms of bilinear operators
\begin{equation}
\hat T^\lambda_\mu=
\sum_{\ell\ell'}t^\lambda_{\ell\ell'}\bigl(b_\ell^\dag\times\tilde b_{\ell'}\bigr)^{(\lambda)}_\mu,
\label{e_slbilin}
\end{equation}
where the summation is over the bosons in the model
and $t^\lambda_{\ell\ell'}$ are pre-defined coefficients.
For example, the quadrupole operator $\hat Q_\mu$ in the SU(3) limit of $sd$-IBM~\cite{Arima78}
is defined with $t^2_{02}=t^2_{20}=1$ and $t^2_{22}=\pm\sqrt{7}/2$.
If several operators occur with the same angular momentum,
they can be distinguished with an additional index $r$ in $t^{\lambda r}_{\ell\ell'}$.

A Hamiltonian with up to $k$-body interactions in the bosons
can be defined in terms of a multipole expansion of the form
\begin{equation}
\bigl(\cdots\bigl(\hat T^{\lambda_1}\times\hat T^{\lambda_2})^{(\Lambda_2)}
\times\hat T^{\lambda_3}\bigr)^{(\Lambda_3)}\times\cdots\times\hat T^{\lambda_k}\bigr)^{(\Lambda_k)},
\label{e_slhmul}
\end{equation}
where for a scalar operator $\Lambda_k=0$.
The bosons in the operator~(\ref{e_slbilin}) can be scalar ($\ell=0$ and/or $\ell'=0$)
and, unlike for an operator in normal-ordered representation,
the $s$-boson dependence of a multipole operator cannot be simply factored out.
It is therefore necessary to assume in this section
that the angular momenta $\ell_i$ can take on arbitrary values, including zero.

The task is therefore to evaluate matrix elements of the operator~(\ref{e_slhmul})
between basis states of $\ell_1\dots\ell_p$-IBM,
which, following the notation of the previous section,
can be written as $|\{\bar\Gamma\bar K\}_p\rangle$,
where $\bar\Gamma$ and $\bar K$ are collective notations
that denote $\{\Gamma_1,\dots,\Gamma_p\}$
and the intermediate angular momenta  $\{K_1,\dots,K_p\}$, respectively,
with $\Gamma_i$ being the labels $\{n_i\upsilon_i\alpha_iJ_i\}$
associated with the $\ell_i$ boson.
The reduced matrix element of an operator in multipole form
between basis states $|\{\bar\Gamma\bar K\}_p\rangle$
can be calculated recursively from
\begin{align}
&\langle\{\bar\Gamma\bar K\}_p\|
\bigl(\cdots\bigl(\hat T^{\lambda_1}\times\hat T^{\lambda_2})^{(\Lambda_2)}
\times\hat T^{\lambda_3}\bigr)^{(\Lambda_3)}\times\cdots\times\hat T^{\lambda_k}\bigr)^{(\Lambda_k)}
\|\{\bar\Gamma'\bar K'\}_p\rangle
\nonumber\\&\quad=
(-)^{K_p+K'_p+\Lambda_k}[\Lambda_k]
\sum_{\{\bar\Gamma''\bar K''\}_p}
\biggl\{\begin{array}{ccc}\Lambda_{k-1}&\lambda_k&\Lambda_k\\K'_p&K_p&K''_p\end{array}\biggr\}
\langle\{\bar\Gamma''\bar K''\}_p\|\hat T^{\lambda_k}\|\{\bar\Gamma'\bar K'\}_p\rangle
\nonumber\\&\qquad\times
\langle\{\bar\Gamma\bar K\}_p\|
\bigl(\cdots\bigl(\hat T^{\lambda_1}\times\hat T^{\lambda_2})^{(\Lambda_2)}
\times\hat T^{\lambda_3}\bigr)^{(\Lambda_3)}\times\cdots\times\hat T^{\lambda_{k-1}}\bigr)^{(\Lambda_{k-1})}
\|\{\bar\Gamma''\bar K''\}_p\rangle,
\label{e_slmem}
\end{align}
where it is assumed that,
in the final application of this recurrence relation for $k=2$, $\Lambda_1=\lambda_1$.
This relation is recurrent in $k$,
the number of bilinear operators in the multipole expansion~(\ref{e_slhmul}),
and all matrix elements in Eq.~(\ref{e_slmem}) are defined in $\ell_1\dots\ell_p$-IBM.
The angular momenta $K_p$, $K'_p$ and $K''_p$
are the final entries in the series $\bar K$, $\bar K'$ and $\bar K''$,
and correspond to the total angular momenta of the bra, ket and intermediate state,
that is, $J=K_p$, $J'=K'_p$ and $J''=K''_p$.
The recurrence relation~(\ref{e_slmem}) shows
that the reduced matrix element of a multipole operator of order $k$
can be converted into a linear combination of products
of $k$ reduced matrix elements of the type
$\langle\{\bar\Gamma\bar K\}_p\|\hat T^{\lambda_i}\|\{\bar\Gamma'\bar K'\}_p\rangle$.
Although the operator $\hat T^{\lambda_i}$
refers to only two bosons with angular momenta $\ell_i$ and $\ell'_i$
(which may be zero),
the states in bra and ket of the matrix element
$\langle\{\bar\Gamma\bar K\}_p\|\hat T^{\lambda_i}\|\{\bar\Gamma'\bar K'\}_p\rangle$
are defined in $\ell_1\dots\ell_p$-IBM.
The latter matrix element can be evaluated with the help of Eq.~(\ref{e_slprmes2})
with the substitutions $B_{\{\bar\Lambda\bar I\}_p}^\dag\rightarrow b_{\ell_i}^\dag$
and $\tilde B_{\{\bar\Lambda'\bar I'\}_p}\rightarrow\tilde b_{\ell'_i}$.

\section{From multipole to normal-ordered interactions}
\label{s_mulnor}
The previous section specified the algorithm to calculate matrix elements in $\ell_1\dots\ell_p$-IBM
of a Hamiltonian in multipole form.
The algorithm is doubly recursive.
First, it reduces a matrix element of a product of $k$ bilinear operators $\hat T^\lambda_\mu$
to a linear combination of matrix elements of products of $k-1$ bilinear operators
and, second, it reduces a matrix element of $\hat T^\lambda_\mu$ in $\ell_1\dots\ell_p$-IBM
to a linear combination of products of a matrix element
in $\ell_1\dots\ell_{p-1}$-IBM times one in $\ell_p$-IBM.
A more efficient algorithm can be defined
by converting the multipole Hamiltonian from the start to its normal-ordered representation.

To achieve the conversion from multipole to normal-ordered form,
one calculates all possible matrix elements in the basis~(\ref{e_slpbasis}), diagonal and off-diagonal,
for boson numbers $N=0,1,\dots,k_{\rm max}$,
where $k_{\rm max}$ is the maximum order of the multipole expansion.
With use of Eq.~(\ref{e_slpmes2}) the matrix elements can expressed analytically
in terms of the $k$-body interactions
$v_{\bar\Lambda\bar I,\bar\Lambda'\bar I'}^{kL}$.
Alternatively, with use of Eq.~(\ref{e_slmem}),
one calculates the same set of matrix elements for the multipole Hamiltonian
and equates them to the corresponding expressions for the normal-ordered Hamiltonian.
The resulting set of linear equations can be solved
and gives the interactions
$v_{\bar\Lambda\bar I,\bar\Lambda'\bar I'}^{kL}$
in terms of the parameters of the multipole Hamiltonian.

This procedure makes use of the generic expressions of the matrix elements
of a Hamiltonian in its normal-ordered and multipole representations
but these are only needed for boson numbers up to and including $k_{\rm max}$.
Once the conversion to the Hamiltonian's normal-ordered representation is obtained,
the latter can be used for an $N$-boson calculation.
Equations~(\ref{e_slprmes2}) and (\ref{e_slmem})
provide entirely different recursive schemes to calculate matrix elements in $\ell_1\dots\ell_p$-IBM
and therefore the conversion from one to the other representation also provides a rigorous test
of the present formalism and its numerical implementation
discussed in the next section.

\section{Numerical implementation}
\label{s_numimp}
Several computer codes are available
to solve the secular equation of a Hamiltonian matrix for a system of interacting bosons.
For applications to nuclei in the framework of the IBM
the first such code, named {\tt phint}, was written by Olaf Scholten~\cite{Scholten79,Scholten80}.
It solves numerically the eigenvalue problem of a Hamiltonian in $sd$-IBM
with single-boson energies and two-boson interactions.
This code has been the basis for numerous extensions.
For example, the version {\tt ibm.f}~\cite{Isacker_ibm.f}
includes all three-body interactions between $s$ and $d$ bosons
and incorporates mixing between configurations
with different total boson number $N$
(for an application, see Ref.~\cite{Garcia-Ramos14}).
A more versatile interacting boson code, named {\tt ArbModel},
was developed by Stefan Heinze~\cite{Heinze08}.
It can deal with bosons with arbitrary angular momentum
and CFPs are calculated as the square-root of rational numbers,
which prevents the loss of precision for large $N$.
The interactions in {\tt ArbModel} are limited to two-body at most
and the Hamiltonian matrix is constructed numerically.

It is worthwhile to compare the available nuclear-physics codes for interacting bosons
with those for fermions, that is, with shell-model codes.
There exists a panoply of the latter.
The most recent ones, still in use, are
{\tt nushellx}~\cite{Brown01,Brown14},
{\tt antoine}~\cite{Caurier05},
{\tt kshell}~\cite{Shimizu12,Shimizu19}
and {\tt bigstick}~\cite{Johnson13,Johnson18}.
They can handle a large number of single-particle orbitals
(corresponding to $p$ in the present formalism)
but are generally restricted to two-body interactions,
which may be made nucleon-number dependent
to approximate microscopically calculated three-nucleon interactions.
Dimensions of the Hamiltonian matrix in the shell model
typically are much larger than those in boson models.
For this reason all current shell-model codes
are written in an $m$-scheme basis,
such that matrix elements must not be stored but can be calculated on the fly.
To the best of the author's knowledge,
there exist no publicly available shell-model code
that is able to express symbolically matrix elements
in a rotationally-invariant $N$-nucleon basis 
in terms of interaction matrix elements of arbitrary order $k$.

The formalism presented in this paper
has been implemented in a Mathematica code named {\tt ibm.m}~\cite{Isacker_ibm.m}.
It accepts $p$ different kinds of bosons with arbitrary angular momenta $\ell_1,\ell_2,\dots,\ell_p$,
with an interaction of arbitrary order $k$.
In addition, the Hamiltonian matrix for $N$ bosons is constructed symbolically,
that is, all its elements are known analytically
as a linear combination of the $k$-body interaction matrix elements.

While the formalism is valid for arbitrary $p$, $k$ and $N$,
it is clear that a numerical implementation imposes limitations on their values.
In all boson models considered to date $p$ is rather low:
the most elaborate model in this respect is the $sdpf$-IBM,
for which $p=3$ since the $s$ boson can be eliminated
before the recursive procedure is started.
Likewise, in all reasonable applications
the order $k$ of the interaction is unlikely to cause numerical problems.
For $sd$-IBM an example with $k_{\rm max}=4$ is discussed in Sect.~\ref{s_exam}
but the order can be further increased to $k=5$ or even $k=6$,
in which case there are 41 and 176 Hermitian interaction matrix elements, respectively.
In more elaborate models, like $sdg$-IBM or $spdf$-IBM,
the number of interaction matrix elements steeply increases with $k$.
For example, in $sdg$-IBM there are 32 two-body Hermitian matrix elements
and this number increases to 324 and 3425 for $k=3$ and $k=4$, respectively.
The corresponding numbers in $spdf$-IBM are 66, 976 and 13038,
if parity conservation is imposed.
While some method is needed
to determine the numerical values of such large sets of interaction matrix elements,
either from microscopic considerations
or from a physics-based expansion of a Hamiltonian in multipole form,
the code {\tt ibm.m} allows the construction of matrix elements in an $N$-boson basis 
for all such values of $k$, provided $N$ is not too large.

It can be concluded therefore that neither $p$ nor $k$
represents a limiting factor in the numerical implementation of the present formalism.
The combination of a high boson angular momentum $\ell$
and a large boson number $N$, however,
does lead to numerical difficulties for the following reason.
The algorithm outlined in Sect.~\ref{s_cfp}
calculates a CFP as the square-root of rational number
and loss of precision is therefore not an issue.
The number of states in the Gram--Schmidt orthonormalisation procedure, however,
constitutes a limitation of the method.
This number is given by the multiplicity $d_\ell(\upsilon,L)$ of Eq.~(\ref{e_slmult}).
For $\ell=2$, as needed in $sd$-IBM,
the number is small for all applications to nuclei;
for example, $d_2(\upsilon,L)\leq4$ for $\upsilon\leq20$ and arbitrary $L$.
The multiplicity increases dramatically with $\ell$;
for example, for a $g$ boson, $d_4(\upsilon,L)$ increases to 50 for $\upsilon=10$
and to 778 for $\upsilon=20$.
The latter number certainly cannot be handled
by the Gram--Schmidt orthonormalisation procedure of Eqs.~(\ref{e_gsseq}--\ref{e_over1})
and therefore an alternative, hitherto unknown, algorithm for the construction of CFPs must be devised.
With the current algorithm calculations in $sdg$-IBM can be carried out
by restricting the number of $g$ bosons to a lower value than the total boson number $N$.

\section{Examples}
\label{s_exam}
\subsection{The classical limit of the $sd$-IBM}
\label{ss_climitsd}
Geometric insight into algebraic models
can be obtained from their classical limit
(i.e., the limit of large boson number $N$)
by means of a coherent state~\cite{Ginocchio80,Dieperink80,Bohr80}.
The topic of this subsection is the classical limit of $sd$-IBM
with the coherent state
\begin{equation}
|N;\beta\gamma\rangle=
\frac{1}{\sqrt{N!(1+\beta^2)^N}}
\biggl(s^\dag+\beta\bigl[\cos\gamma d^\dag_0
+\sqrt{\textstyle{\frac12}}\sin\gamma(d^\dag_{-2}+d^\dag_{+2})\bigr]
\biggr)^N|{\rm o}\rangle,
\label{e_sdcoh1}
\end{equation}
where $|{\rm o}\rangle$ is the boson vacuum
and $\beta$ and $\gamma$ are the quadrupole shape parameters.
The calculation of the Hamiltonian's expectation value in this state
leads to a function of $N$, $\beta$ and $\gamma$,
which has the interpretation of an energy surface in the shape variables
and depends on the interaction matrix elements $v_{n\upsilon,n'\upsilon'}^{kL}$
defined in Eq.~(\ref{e_slkbody}).
More generally, the coherent-state formalism
can also be used to calculate off-diagonal matrix elements,
which have the interpretation of interactions between intrinsic states.
This requires the generalised definition
\begin{equation}
|N;\bar c\rangle=
\frac{1}{\sqrt{N!(1+c^2)^N}}
(B^\dag_c)^N|{\rm o}\rangle,
\quad
B^\dag_c=s^\dag+\sum_{\mu=-2}^{+2}c_\mu d_\mu^\dag,
\label{e_sdcoh2}
\end{equation}
where $c^2=\sum_\mu c_\mu^*c_\mu$.
With $c_0=\beta\cos\gamma$, $c_{\pm1}=0$ and $c_{\pm2}=\frac{1}{\sqrt2}\beta\sin\gamma$,
the coherent state~(\ref{e_sdcoh1}) is obtained
but the definition~(\ref{e_sdcoh2}) covers other cases as needed,
for example, in the calculation of the matrix element
between the coherent state~(\ref{e_sdcoh1}) and the one obtained after rotation.
The generalised matrix element is calculated to be
\begin{align}
\langle N;\bar c|(-)^LB_{kn\upsilon L}^\dag\cdot\tilde B_{kn'\upsilon'L}|N;\bar c'\rangle={}&
\frac{N!}{(N-k)!}
\frac{(1+c^*\cdot c')^{N-k}}{\sqrt{(1+c^2)^N(1+c'^2)^N}}
\frac{1}{\sqrt{(k-m)!(k-m')!}}
\nonumber\\&\times
\sum_M\sum_{\{\mu_i\}}\sum_{\{\mu_{i'}\}}
a_{n\upsilon LM}^{\{\mu_i\}*}a_{n'\upsilon'LM}^{\{\mu_{i'}\}}
\prod_{i=1}^mc_{\mu_i}^*\prod_{i'=1}^{m'}c'_{\mu_{i'}}.
\label{e_sdcohoff}
\end{align}
For a diagonal matrix element, $\bar c=\bar c'$, this expression simplifies to
\begin{align}
&\langle N;\bar c|(-)^LB_{kn\upsilon L}^\dag\cdot\tilde B_{kn'\upsilon'L}|N;\bar c\rangle
\nonumber\\&\quad=
\frac{N!}{(N-k)!}
\frac{1}{(1+c^2)^k}
\frac{1}{\sqrt{(k-m)!(k-m')!}}
\sum_M\sum_{\{\mu_i\}}\sum_{\{\mu_{i'}\}}
a_{n\upsilon LM}^{\{\mu_i\}*}a_{n'\upsilon'LM}^{\{\mu_{i'}\}}
\prod_{i=1}^mc_{\mu_i}^*\prod_{i'=1}^{m'}c_{\mu_{i'}}.
\label{e_sdcohdia}
\end{align}
Use is made of the expansion of $B_{n\upsilon LM}^\dag$
in terms of uncoupled $d$-boson creation operators,
\begin{equation}
B_{n\upsilon LM}^\dag=
\sum_{\{\mu_i\}}a_{n\upsilon LM}^{\{\mu_i\}}
d_{\mu_1}^\dag d_{\mu_2}^\dag\dots d_{\mu_n}^\dag,
\label{e_slBexp}
\end{equation}
where $\{\mu_i\}$ refers to all possible sets $\{\mu_1,\mu_2,\dots,\mu_n\}$
such that $\mu_1+\mu_2+\cdots+\mu_n=M$.
The expectation value can then be evaluated
by noting that the action of the uncoupled creation and annihilation operators
is equivalent to a derivative~\cite{Isacker81}.

The coefficients $a_{n\upsilon LM}^{\{\mu_i\}}$ can be found
by diagonalising a $d$-boson Hamiltonian with up to two-body interactions in the $m$ scheme.
The quantum numbers $n$, $\upsilon$ and $L$ of the eigenstates for a given value of $M$
can be obtained from the eigenenergies since these are known analytically.
The corresponding eigenvectors determine the coefficients $a_{n\upsilon LM}^{\{\mu_i\}}$.
A difficulty arises as the signs of the coefficients with fixed $n$, $\upsilon$ and $L$ but varying $M$
are related through a Clebsch-Gordan series whereas numerically they are not.
To circumvent this problem,
one calculates the coefficient $a_{n\upsilon LM}^{\{\mu_i\}}$ for a single value of $M$, e.g., $M=-L$,
and subsequently applies the raising operator $\hat L_+$
with use of the property $\hat L_+|LM\rangle=\sqrt{(L-M)(L+M+1)}\hat L_+|L,M+1\rangle$.
To obtain a normalised state, one uses the operator
\begin{equation}
\hat T_+=
\frac{\hat L_+}{\sqrt{(L-M)(L+M+1)}}=
\sum_{m=-\ell}^{\ell-1}r_m b_{\ell,m+1}^\dag b_{\ell m},
\label{e_slraising}
\end{equation}
where
\begin{equation}
r_m=(-)^{\ell+m+1}
\left[\frac{2\ell(\ell+1)(2\ell+1)}{3(L-M)(L+M+1)}\right]^{1/2}
(\ell m+1\;\ell-m|1+1),
\label{e_slrcoef}
\end{equation}
with $\ell=2$ in the case of $d$ bosons.

The algorithm outlined above can be used
to determine the classical limit of a Hamiltonian in $sd$-IBM
with up to and including five-body interactions.
For interactions of higher order than five,
the quantum numbers of seniority and angular momentum
no longer suffice to uniquely characterise an operator $B_{n\upsilon LM}^\dag$
and an additional label $\alpha$ is sometimes needed.
For example, for $\upsilon=6$ two states occur with angular momentum $L=6$,
\mbox{$d_2(\upsilon=L=6)=2$}.
A possible strategy to deal with such even more complicated interactions
is to define the coefficients $a_{n\upsilon LM}^{\{\mu_i\}}$
by diagonalising in the $m$ scheme a $d$-boson Hamiltonian that includes a three-body interaction.
The most logical choice is to consider $v_{ddd\cdot ddd}^0$.
This particular interaction is expected to be related to $n_\Delta$,
which is usually introduced as an additional label~\cite{Iachello87}
and is associated with the number of triplets of $d$ bosons coupled to angular momentum zero
(see also the discussion in chapter~8 of Ref.~\cite{Frank94}).
In fact, it can be shown that the operator $\tilde B_{n=\upsilon=3,L=M=0}$
annihilates a particular combination of the two states with $n=\upsilon=6$ and $L=6$,
and this provides a possible definition of the state $|n=\upsilon=6,\alpha=1,L=6\rangle$
and therefore of the coefficients $a_{n\upsilon LM}^{\{\mu_i\}}$.

The energy surface of a generic $sd$-IBM Hamiltonian with interactions up to order $k_{\rm max}$
can be written in the following manner:
\begin{equation}
\langle N;\beta\gamma|\hat H_{\rm IBM}|N;\beta\gamma\rangle=
E_0+\sum_{k=1}^{k_{\rm max}}\frac{N(N-1)\cdots(N-k+1)}{(1+\beta^2)^k}
\sum_{rt}a^{(k)}_{rt}\beta^{2r+3t}(\cos3\gamma)^t,
\label{e_sdclimit}
\end{equation}
where the second summation is over all non-negative integer values of $r$ and $t$ such that $2r+3t\leq2k$.
The coefficients $a^{(k)}_{rt}$ are well known (see, e.g., Ref.~\cite{Isacker81}) for $k=1$,
\begin{equation}
a^{(1)}_{00}=\epsilon_s,
\quad
a^{(1)}_{10}=\epsilon_d,
\label{e_sdclimit1}
\end{equation}
where $\epsilon_s$ and $\epsilon_d$ are the $s$- and $d$-boson energies,
and for $k=2$,
\begin{align}
&\textstyle
a^{(2)}_{00}={\frac12}v_{ss\cdot ss}^0,
\quad
a^{(2)}_{10}=
\frac{1}{\sqrt 5}v_{ss\cdot dd}^0
+v_{sd\cdot sd}^2,
\quad
a^{(2)}_{01}=-{\frac{2}{\sqrt 7}}v_{sd\cdot dd}^2,
\nonumber\\&\textstyle
a^{(2)}_{20}=
\frac{1}{10}v_{dd\cdot dd}^0+{\frac17}v_{dd\cdot dd}^2
+\frac{9}{35}v_{dd\cdot dd}^4,
\label{e_sdclimit2}
\end{align}
which are written in terms of the combinations
\begin{equation}
v_{kn\upsilon\cdot kn'\upsilon'}^L=
{\frac12}\bigl(v_{kn\upsilon,kn'\upsilon'}^L+v_{kn'\upsilon',kn\upsilon}^L\bigr)=
v_{kn\upsilon,kn'\upsilon'}^L,
\label{e_sdinther}
\end{equation}
where the last equality is valid for a Hermitian interaction.

The algorithm described in this section
allows for the systematic calculation of the coefficients $a^{(k)}_{rt}$ for $k>2$
and is part of the code {\tt ibm.m}~\cite{Isacker_ibm.m}.
The coefficients for $k=3$ and 4 are given in the Appendix.
From this extended analysis the following conclusions can be drawn.
The number of $d$ bosons involved in the interaction
determines the power $2r+3t$ of $\beta$ in the energy surface~(\ref{e_sdclimit}),
that is, $2r+3t=n+n'$
where $n$ and $n'$ refer to the $d$-boson numbers in the interaction~(\ref{e_sdinther}).
A non-zero power of $\cos3\gamma$ introduces a $\gamma$ dependence in the energy surface,
which can acquire a triaxial minimum with $0^{\rm o}<\gamma<60^{\rm o}$
only if $t\geq2$.
A power $t$ of $\cos3\gamma$ requires at least $3t$ $d$ bosons.
In fact, analysis shows that the condition is stronger
since it requires at least $3t$ $d$ bosons
not in pairs coupled to angular momentum $L=0$,
that is, $\lfloor(\upsilon+\upsilon')/3\rfloor\geq t$
where $\upsilon$ and $\upsilon'$ refer to the seniorities in the interaction~(\ref{e_sdinther})
and $\lfloor x\rfloor$ in the largest integer smaller than or equal to $x$.
The explicit formula for $t$ is
\begin{equation}
t_{\rm max}={\frac13}\Bigl(\upsilon+\upsilon'-{\rm mod}_3(\upsilon+\upsilon')
\bigl[7-3\,{\rm mod}_3(\upsilon+\upsilon')\bigr]\Bigr),
\quad
t=t_{\rm max},t_{\rm max}-2,\dots\geq0.
\label{e_sdtform}
\end{equation}
To understand the implications of this formula, one distinguishes the following three cases:
\begin{enumerate}
\item
If ${\rm mod}_3(\upsilon+\upsilon')=0$, then $t_{\rm max}=(\upsilon+\upsilon')/3$.
This proves that a triaxial shape associated with $(\cos3\gamma)^2$
requires at least cubic interactions between the $d$ bosons~\cite{Isacker81}.
Since the $3d$-boson states in $v_{ddd\cdot ddd}^2$ have seniority $\upsilon=1$,
this interaction does not contribute to $(\cos3\gamma)^2$,
in contrast to the interactions $v_{ddd\cdot ddd}^L$ with $L\neq2$,
which have $\upsilon+\upsilon'=6$.
\item
If ${\rm mod}_3(\upsilon+\upsilon')=1$, then $t_{\rm max}=(\upsilon+\upsilon'-4)/3$.
As a consequence, quadratic interactions between the $d$ bosons
are independent of $\gamma$ in the classical limit.
Similarly, since the sum of the  seniorities
in $v_{dddd_2\cdot dddd_2}^2$ or $v_{dddd_2\cdot dddd_2}^4$
is $\upsilon+\upsilon'=4$,
these four-body interactions are also independent of $\gamma$ in the classical limit.
\item
If ${\rm mod}_3(\upsilon+\upsilon')=2$, then $t_{\rm max}=(\upsilon+\upsilon'-2)/3$.
Quartic interactions between the $d$ bosons can have at most $\upsilon+\upsilon'=8$,
in which case $t_{\rm max}=2$,
yielding a term in $(\cos3\gamma)^2$ ($t=2$) and one independent of $\gamma$ ($t=0$).
\end{enumerate}

\subsection{The $\hat Q_\mu+\hat L_\mu$ expansion of the $sd$-Hamiltonian}
\label{ss_QLsd}
Given the importance of quadrupole deformation in nuclei,
several authors have explored the results of an expansion
in the quadrupole operator
\begin{equation}
\hat Q_\mu=(s^\dag\times\tilde d+d^\dag\times\tilde s)^{(2)}_\mu+\chi(d^\dag\times\tilde d)^{(2)}_\mu,
\label{e_qsd}
\end{equation}
with the goal to introduce higher-order interactions in the $sd$-IBM,
see for example Fortunato {\it et al.}~\cite{Fortunato11}.
The scope of such an approach can be enlarged
by including the angular momentum operator $\hat L_\mu$ in the expansion,
in which case eight third-order scalar operators can be constructed, namely
\begin{align}
&(\hat Q\times\hat Q\times\hat Q)^{(0)},
\nonumber\\
&(\hat L\times\hat Q\times\hat Q)^{(0)},
\quad
(\hat Q\times\hat L\times\hat Q)^{(0)},
\quad
(\hat Q\times\hat Q\times\hat L)^{(0)},
\nonumber\\
&(\hat L\times\hat L\times\hat Q)^{(0)},
\quad
(\hat L\times\hat Q\times\hat L)^{(0)},
\quad
(\hat Q\times\hat L\times\hat L)^{(0)},
\quad
\nonumber\\
&(\hat L\times\hat L\times\hat L)^{(0)},
\label{e_ql3a}
\end{align}
where in each term the first two operators
are coupled to the angular momentum of the third operator
to yield an overall scalar.

\begin{table}
\centering
\caption{The single-boson energies and the normal-ordered two-boson interactions
for the eight cubic operators of Eq.~(\ref{e_ql3a}).}
\label{t_ql3a}
\begin{ruledtabular}
\begin{tabular}{l|ccccc}
&$(\hat Q\!\times\!\hat Q\!\times\!\hat Q)^{(0)}$&
$(\hat Q\!\times\!\hat L\!\times\!\hat Q)^{(0)}$&
$(\hat L\!\times\!\hat Q\!\times\!\hat Q)^{(0)}$&
$(\hat L\!\times\!\hat L\!\times\!\hat Q)^{(0)}$&
$(\hat L\!\times\!\hat L\!\times\!\hat L)^{(0)}$\\
&&&$(\hat Q\!\times\!\hat Q\!\times\!\hat L)^{(0)}$&
$(\hat L\!\times\!\hat Q\!\times\!\hat L)^{(0)}$&\\
&&&&
$(\hat Q\!\times\!\hat L\!\times\!\hat L)^{(0)}$&\\
\hline
$\epsilon_s$&$\sqrt{5}\chi$&$-\sqrt{30}$&0&0&0\\
$\epsilon_d$&$\frac{\chi(28-3\chi^2)}{14\sqrt{5}}$&
$-{\frac{\sqrt{3}\chi^2}{\sqrt{10}}}$&
$-\frac{\sqrt{3}(2+\chi^2)}{\sqrt{10}}$&
$\frac{\sqrt{21}\chi}{\sqrt{5}}$&$\sqrt{6}$\\
\hline
$v_{ss\cdot ss}^0$&0&0&0&0&0\\
$v_{ss\cdot dd}^0$&$6\chi$&$-2\sqrt{6}$&0&0&0\\
$v_{dd\cdot dd}^0$&$\frac{3\chi(14-3\chi^2)}{7\sqrt{5}}$&
$-\frac{\sqrt{6}(2+3\chi^2)}{\sqrt{5}}$&
$\frac{\sqrt{6}(2+\chi^2)}{\sqrt{5}}$&
$-\frac{2\sqrt{21}\chi}{\sqrt{5}}$&$-2\sqrt{6}$\\
$v_{sd\cdot sd}^2$&$\frac{6\chi}{\sqrt{5}}$&$-\frac{2\sqrt{6}}{\sqrt{5}}$&0&0&0\\
$v_{sd\cdot dd}^2$&$\frac{3(14+11\chi^2)}{7\sqrt{10}}$&$-\frac{4\sqrt{3}\chi}{\sqrt{5}}$&0&$\sqrt{\frac{42}{5}}$&0\\
$v_{dd\cdot dd}^2$&$\frac{9\chi(-14+3\chi^2)}{98\sqrt{5}}$&
$-\frac{\sqrt{3}(14+\chi^2)}{7\sqrt{10}}$&
$\frac{\sqrt{3}(2+\chi^2)}{\sqrt{10}}$&
$-\frac{17\sqrt{3}\chi}{\sqrt{35}}$&$-\sqrt{6}$\\
$v_{dd\cdot dd}^4$&$\frac{6\chi(14-3\chi^2)}{49\sqrt{5}}$&
$\frac{2\sqrt{2}(14+\chi^2)}{7\sqrt{15}}$&
$-\frac{2\sqrt{2}(2+\chi^2)}{\sqrt{15}}$&
$\frac{68\chi}{\sqrt{105}}$&$4\sqrt{\frac{2}{3}}$\\
\end{tabular}
\end{ruledtabular}
\end{table}
\begin{table}
\caption{The normal-ordered three-boson interactions
for four of the eight cubic operators of Eq.~(\ref{e_ql3a}).}
\label{t_ql3b}
\begin{ruledtabular}
\begin{tabular}{l|cc||l|cc}
&$(\hat Q\!\times\!\hat Q\!\times\!\hat Q)^{(0)}$&$(\hat L\!\times\!\hat L\!\times\!\hat Q)^{(0)}$&
&$(\hat Q\!\times\!\hat Q\!\times\!\hat Q)^{(0)}$&$(\hat L\!\times\!\hat L\!\times\!\hat Q)^{(0)}$\\
&&$(\hat L\!\times\!\hat Q\!\times\!\hat L)^{(0)}$&&&$(\hat L\!\times\!\hat Q\!\times\!\hat L)^{(0)}$\\
&&$(\hat Q\!\times\!\hat L\!\times\!\hat L)^{(0)}$&&&$(\hat Q\!\times\!\hat L\!\times\!\hat L)^{(0)}$\\
\hline
$v_{sss\cdot sss}^0$&0&0&
$v_{ssd\cdot ssd}^2$&0&0\\
$v_{sss\cdot sdd}^0$&0&0&
$v_{ssd\cdot sdd}^2$&$\frac{6}{\sqrt{5}}$&0\\
$v_{sss\cdot ddd}^0$&6&0&
$v_{ssd\cdot ddd}^2$&$\frac{12\chi}{\sqrt{35}}$&0\\
$v_{sdd\cdot sdd}^0$&$\frac{12\chi}{\sqrt{5}}$&0&
$v_{sdd\cdot sdd}^2$&$-\frac{18\chi}{7\sqrt{5}}$&0\\
$v_{sdd\cdot ddd}^0$&$-\frac{9\sqrt{3}\chi^2}{7\sqrt{5}}$&$-6\sqrt{\frac{7}{5}}$&
$v_{sdd\cdot ddd}^2$&$\frac{15\sqrt{5}\chi^2}{7\sqrt{7}}$&$-2\sqrt{\frac{3}{5}}$\\
$v_{ddd\cdot ddd}^0$&$\frac{123\chi^3}{49\sqrt{5}}$&$\frac{6\sqrt{15}\chi}{\sqrt{7}}$&
$v_{ddd\cdot ddd}^2$&$-\frac{18\chi^3}{49\sqrt{5}}$&$-\frac{4\sqrt{3}\chi}{\sqrt{35}}$\\
$v_{sdd\cdot sdd}^4$&$\frac{24\chi}{7\sqrt{5}}$&0&
$v_{ddd\cdot ddd}^3$&$-\frac{114\chi^3}{49\sqrt{5}}$&$-\frac{4\sqrt{3}\chi}{\sqrt{35}}$\\
$v_{sdd\cdot ddd}^4$&$\frac{6\sqrt{11}\chi^2}{7\sqrt{35}}$&$4\sqrt{\frac{11}{15}}$&
$v_{ddd\cdot ddd}^6$&$-\frac{24\chi^3}{49\sqrt{5}}$&$\frac{16\sqrt{3}\chi}{\sqrt{35}}$\\
$v_{ddd\cdot ddd}^4$&$\frac{108\chi^3}{49\sqrt{5}}$&$-\frac{8\sqrt{5}\chi}{\sqrt{21}}$&&&\\
\end{tabular}
\end{ruledtabular}
\end{table}
To answer the question how the operators~(\ref{e_ql3a}) are related to each other,
one applies the algorithm of Sect.~\ref{s_mulnor}
that converts them to a normal-ordered representation.
The boson energies and the two-boson interactions
of the eight operators are shown in Table~\ref{t_ql3a}.
The normal-ordered representation shows that
\begin{equation}
(\hat L\times\hat Q\times\hat Q)^{(0)}=
(\hat Q\times\hat Q\times\hat L)^{(0)}\neq
(\hat Q\times\hat L\times\hat Q)^{(0)},
\label{e_ql3b}
\end{equation}
and that the three-body part of these three operators vanishes for any value of $\chi$.
Similarly, the three-boson interaction of $(\hat L\times\hat L\times\hat L)^{(0)}$ vanishes identically.
The only operators in the list~(\ref{e_ql3a}) with a non-zero three-boson interaction
are those with three $\hat Q_\mu$ operators
or with one $\hat Q_\mu$ and two $\hat L_\mu$ operators, see Table~\ref{t_ql3b}.
In addition, the normal-ordered representation shows that
the latter operators are independent of the order of $\hat L_\mu$ and $\hat Q_\mu$,
\begin{equation}
(\hat L\times\hat L\times\hat Q)^{(0)}=
(\hat L\times\hat Q\times\hat L)^{(0)}=
(\hat Q\times\hat L\times\hat L)^{(0)}.
\label{e_ql3c}
\end{equation}
One concludes therefore that there are five independent operators in the list~(\ref{e_ql3a}),
two of which have a non-zero interaction of order $k=3$.

Matrix elements of an operator in multipole form
can be calculated in two different ways,
either with the recursive formula~(\ref{e_slmem})
or by converting to the operator's normal-ordered representation
and applying Eq.~(\ref{e_slpmes2}).
Both methods must yield the same result
but, as remarked earlier, the latter algorithm is more efficient.
This can be illustrated with the example
of the $(\hat Q\times\hat Q\times\hat Q)^{(0)}$ operator in the $sd$-IBM.
For $N=10$ bosons and angular momentum $J=0$
the construction of the entire Hamiltonian matrix is about five times faster
after conversion to normal order.
This factor increases to $\sim$100 for $N=20$ and $J=0$.

\begin{figure}
\centering
\includegraphics[width=6.5cm]{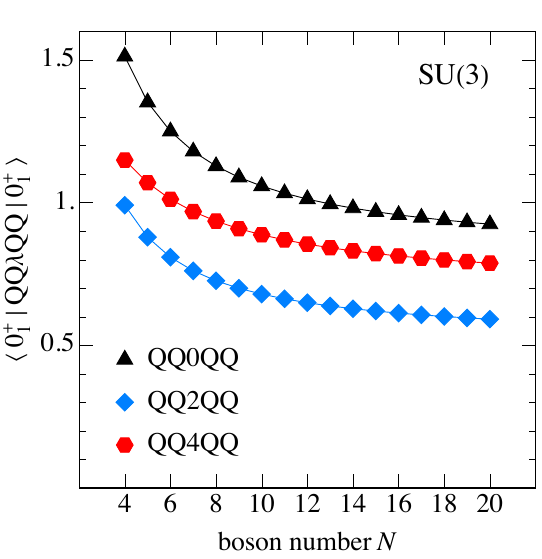}
\includegraphics[width=6.5cm]{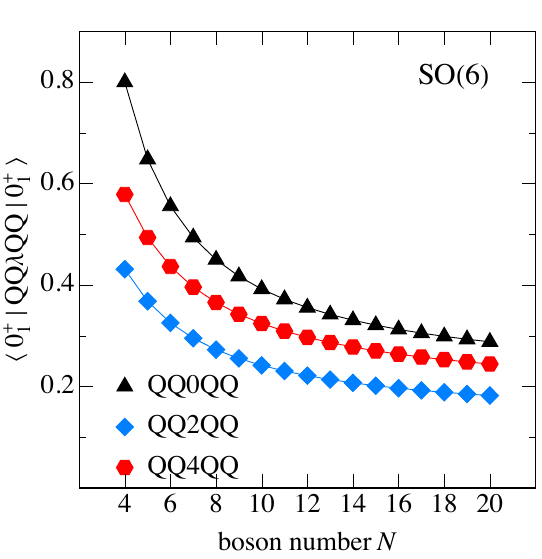}
\caption{The expectation value of the fourth-order operators~(\ref{e_q4a})
in the ground state $|0_1^+\rangle$ of the Hamiltonian $-\kappa\hat Q\cdot\hat Q$
in the SU(3) limit ($\chi=\pm\sqrt{7}/2$) and the SO(6) limit ($\chi=0$).
Results are shown for $\lambda=0$, 2 and 4.
The notation $QQ\lambda QQ$ stands for
$\bigl((\hat Q\times\hat Q)^{(\lambda)}\times(\hat Q\times\hat Q)^{(\lambda)}\bigr)^{(0)}/N^4$.}
\label{f_q4}
\end{figure}
One can extend this analysis to fourth-order multipole operators.
If one considers for simplicity's sake only combinations of $\hat Q_\mu$,
then there are five different operators,
\begin{equation}
\bigl(\bigl((\hat Q\times\hat Q)^{(\lambda)}\times\hat Q\bigr)^{(2)}\times\hat Q\bigr)^{(0)}=
\bigl((\hat Q\times\hat Q)^{(\lambda)}\times(\hat Q\times\hat Q)^{(\lambda)}\bigr)^{(0)},
\quad
\lambda=0,1,2,3,4.
\label{e_q4a}
\end{equation}
How the operators~(\ref{e_q4a}) are related to each other
again can be studied by converting to a normal-ordered representation.
The full details of this analysis will not be given here
but its results can be summarised as follows.
One finds that none of the operators in the set~(\ref{e_q4a}) is equivalent to another one.
The operators~(\ref{e_q4a}) with $\lambda=1$ and 3
have vanishing three- and four-boson interactions,
and their non-zero single-boson energies and two-boson interactions can be written as follows:
\begin{align}
\lambda=1&:\quad\epsilon_d=-3f(\chi),
\quad
v_{dd\cdot dd}^0=6f(\chi),
\quad
v_{dd\cdot dd}^2=3f(\chi),
\quad
v_{dd\cdot dd}^2=-4f(\chi),
\nonumber\\
\lambda=3&:\quad\epsilon_d=-7g(\chi),
\quad
v_{dd\cdot dd}^0=14g(\chi),
\quad
v_{dd\cdot dd}^2=-8g(\chi),
\quad
v_{dd\cdot dd}^2=-g(\chi),
\label{e_q4b}
\end{align}
with
\begin{equation}
f(\chi)=\frac{(2+\chi^2)^2}{20\sqrt{3}},
\quad
g(\chi)=\frac{(7-4\chi^2)^2}{245\sqrt{7}}.
\label{e_q4c}
\end{equation}

The non-equivalence of the five operators~(\ref{e_q4a}) can be illustrated
with the example of their expectation value in a $|0_1^+\rangle$ ground state.
For $\lambda=0$ this expectation value has an important physics interpretation
since it determines fluctuations in the quadrupole deformation parameter $\beta$~\cite{Poves20}.
By way of example one can choose $|0_1^+\rangle$
to be the ground state of $\hat H=-\kappa\hat Q\cdot\hat Q$ with $\kappa>0$,
which is a frequently used $sd$-IBM Hamiltonian
in terms of a single parameter $\chi$~\cite{Warner83}.
For $\lambda=1$ and 3 one finds that the expectation value vanishes identically,
\begin{equation}
\langle0_1^+|\bigl((\hat Q\times\hat Q)^{(1)}\times(\hat Q\times\hat Q)^{(1)}\bigr)^{(0)}|0_1^+\rangle=
\langle0_1^+|\bigl((\hat Q\times\hat Q)^{(3)}\times(\hat Q\times\hat Q)^{(3)}\bigr)^{(0)}|0_1^+\rangle=0,
\label{e_q4d}
\end{equation}
and this is valid for arbitrary values of $\chi$.
The first of these identities is readily understood
since the normal-ordered representation of the $\lambda=1$ operator
given in Eqs.~(\ref{e_q4b}) and~(\ref{e_q4c}) shows that
\begin{equation}
\bigl((\hat Q\times\hat Q)^{(1)}\times(\hat Q\times\hat Q)^{(1)}\bigr)^{(0)}=
-\frac{(2+\chi^2)^2}{40\sqrt{3}}\hat L^2.
\label{e_q4e}
\end{equation}
In the three remaining cases with $\lambda=0$, 2 and 4 the expectation value
\begin{equation}
\frac{1}{N^4}
\langle0_1^+|\bigl((\hat Q\times\hat Q)^{(\lambda)}\times(\hat Q\times\hat Q)^{(\lambda)}\bigr)^{(0)}|0_1^+\rangle,
\label{e_q4f}
\end{equation}
is shown in Fig.~\ref{f_q4} for two values of $\chi$:
its SU(3) value $\chi=\pm\sqrt{7}/2$ and its SO(6) value $\chi=0$.
One observes a clear dependence on $\lambda$ for all $N$
and this proves the non-equivalence of the five $k=4$ operators~(\ref{e_q4a}).

\section{Concluding remarks}
\label{s_conc}
The formalism presented in this paper
enables the solution of the eigenvalue problem
for a rotationally invariant and boson-number conserving Hamiltonian in $\ell_1\dots\ell_p$-IBM.
In addition, matrix elements can be evaluated of tensor operators
that are not necessarily scalar nor boson-number conserving.
There is, in principle, no restriction
on $N$, the number of bosons,
on $p$, the number of different kinds of bosons,
and on $k$, the order of the interaction between the bosons
but the overall computational cost mounts with increasing $N$, $p$ and $k$.
Specifically, the numerical implementation of the formalism is mostly limited
by a combination of a high boson number $N$ with a high boson angular momentum $\ell$.

One of the characteristic advantages of the symbolic method is
that it expresses a Hamiltonian matrix element between $N$-boson states
as a linear combination of the interaction matrix elements.
This feature may give insight into the structure of the Hamiltonian matrix.
To obtain the eigenspectrum and eigenfunctions, this matrix must be diagonalised.
With very rare exceptions, this can only be achieved
by inserting appropriate numerical values for the interaction matrix elements.

It is relatively straightforward to extend the present formalism to a system of interacting fermions,
with possible applications to the nuclear shell model.
Work in this direction is in progress.
However, dimensions can be very much larger in the shell model than they are in the IBM,
in which case it is more efficient to revert
to a standard approach~\cite{Brown01,Brown14,Caurier05,Shimizu12,Shimizu19,Johnson13,Johnson18}.
If shell-model dimensions are not too large,
the symbolic method may prove to be useful
because it reveals the analytic dependence of the many-body Hamiltonian matrix
on the interaction matrix elements.

\section*{Appendix}
\label{s_app}
This appendix lists the coefficients $a^{(k)}_{rt}$ in Eq.~(\ref{e_sdclimit}) for $k=3$ and 4.
They are expressed in terms of the interaction matrix elements~(\ref{e_sdinther})
and have been obtained with the code {\tt ibm.m}~\cite{Isacker_ibm.m}.
For $k=3$ the coefficients $a^{(3)}_{rt}$ have been derived previously~\cite{Sorgunlu08}
and are repeated here for completeness,
\begin{align}
&\textstyle
a^{(3)}_{00}={\frac16}v_{sss\cdot sss}^0,
\quad
a^{(3)}_{10}=
\frac{1}{\sqrt{15}}v_{sss\cdot sdd}^0
+{\frac12}v_{ssd\cdot ssd}^2,
\nonumber\\&\textstyle
a^{(3)}_{01}=-{\frac13}\sqrt{\frac{2}{35}}v_{sss\cdot ddd}^0
-\sqrt{\frac27}v_{ssd\cdot sdd}^2,
\nonumber\\&\textstyle
a^{(3)}_{20}=
\frac{1}{10}v_{sdd\cdot sdd}^0
+{\frac17}v_{sdd\cdot sdd}^2
+\frac{1}{\sqrt 7}v_{ssd\cdot ddd}^2
+\frac{9}{35}v_{sdd\cdot sdd}^4,
\nonumber\\&\textstyle
a^{(3)}_{11}=
-{\frac15}\sqrt{\frac{2}{21}}v_{sdd\cdot ddd}^0
-{\frac{\sqrt 2}{7}}v_{sdd\cdot ddd}^2
-\frac{18}{35}\sqrt{\frac{2}{11}}v_{sdd\cdot ddd}^4,
\nonumber\\&\textstyle
a^{(3)}_{30}=
\frac{1}{14}v_{ddd\cdot ddd}^2
+\frac{1}{30}v_{ddd\cdot ddd}^3
+\frac{3}{154}v_{ddd\cdot ddd}^4
+\frac{7}{165}v_{ddd\cdot ddd}^6,
\nonumber\\&\textstyle
a^{(3)}_{02}=
\frac{1}{105}v_{ddd\cdot ddd}^0
-\frac{1}{30}v_{ddd\cdot ddd}^3
+\frac{3}{110}v_{ddd\cdot ddd}^4
-\frac{4}{1155}v_{ddd\cdot ddd}^6.
\label{e_sdclimit3}
\end{align}
For $k=4$ the coefficients $a^{(4)}_{rt}$ are
\begin{align}
&\textstyle
a^{(4)}_{00}=\frac{1}{24}v_{ssss\cdot ssss}^0,
\quad
a^{(4)}_{10}=
\frac{1}{2\sqrt{30}}v_{ssss\cdot ssdd}^0
+{\frac16}v_{sssd\cdot sssd}^2,
\nonumber\\&\textstyle
a^{(4)}_{01}=
-\frac{1}{3\sqrt{70}}v_{ssss\cdot sddd}^0
-\frac{1}{\sqrt{21}}v_{sssd\cdot ssdd}^2,
\nonumber\\&\textstyle
a^{(4)}_{20}=
\frac{1}{20}v_{ssdd\cdot ssdd}^0
+\frac{1}{4\sqrt{105}}v_{ssss\cdot dddd}^0
+\frac{1}{14}v_{ssdd\cdot ssdd}^2
+\frac{1}{\sqrt{21}}v_{sssd\cdot sddd}^2
+\frac{9}{70}v_{ssdd\cdot ssdd}^4,
\nonumber\\&\textstyle
a^{(4)}_{11}=
-\frac{1}{5\sqrt{21}}v_{ssdd\cdot sddd}^0
-{\frac17}v_{ssdd\cdot sddd}^2
-\frac{1}{3\sqrt{21}}v_{sssd\cdot dddd_2}^2
-\frac{2}{3\sqrt{231}}v_{sssd\cdot dddd_4}^2
-\frac{18}{35\sqrt{11}}v_{ssdd\cdot sddd}^4,
\nonumber\\&\textstyle
a^{(4)}_{30}=
\frac{1}{14}v_{sddd\cdot sddd}^2
+\frac{1}{30}v_{sddd\cdot sddd}^3
+\frac{3}{154}v_{sddd\cdot sddd}^4
+\frac{7}{165}v_{sddd\cdot sddd}^6
\nonumber\\&\textstyle\phantom{a^{(4)}_{30}}
+\frac{1}{10\sqrt{14}}v_{ssdd\cdot dddd}^0
+\frac{1}{21}v_{ssdd\cdot dddd_2}^2
-\frac{1}{21\sqrt{11}}v_{ssdd\cdot dddd_4}^2
+\frac{3}{35}v_{ssdd\cdot dddd_2}^4
+\frac{3}{7\sqrt{1430}}v_{ssdd\cdot dddd_4}^4,
\nonumber\\&\textstyle
a^{(4)}_{02}=
\frac{1}{105}v_{sddd\cdot sddd}^0
-\frac{1}{30}v_{sddd\cdot sddd}^3
+\frac{3}{110}v_{sddd\cdot sddd}^4
-\frac{4}{1155}v_{sddd\cdot sddd}^6
\nonumber\\&\textstyle\phantom{a^{(4)}_{02}}
+\frac{1}{7\sqrt{11}}v_{ssdd\cdot dddd_4}^2
-\frac{9}{7\sqrt{1430}}v_{ssdd\cdot dddd_4}^4,
\nonumber\\&\textstyle
a^{(4)}_{21}=
-\frac{1}{35\sqrt{6}}v_{sddd\cdot dddd}^0
-\frac{1}{21}v_{sddd\cdot dddd_2}^2
-\frac{2}{21\sqrt{11}}v_{sddd\cdot dddd_4}^2
-\frac{6}{35\sqrt{11}}v_{sddd\cdot dddd_2}^4
\nonumber\\&\textstyle\phantom{a^{(4)}_{21}}
+\frac{6}{77}\sqrt{\frac{2}{65}}v_{sddd\cdot dddd_4}^4
-\frac{6}{77{\sqrt5}}v_{sddd\cdot dddd}^6,
\nonumber\\&\textstyle
a^{(4)}_{40}=
\frac{1}{280}v_{dddd\cdot dddd}^0
+\frac{1}{126}v_{dddd_2\cdot dddd_2}^2
+\frac{1}{1386}v_{dddd_4\cdot dddd_4}^2
+\frac{1}{70}v_{dddd_2\cdot dddd_2}^4
+\frac{17}{5005}v_{dddd_4\cdot dddd_4}^4
\nonumber\\&\textstyle\phantom{a^{(4)}_{40}}
+\frac{1}{210}v_{dddd\cdot dddd}^5
+\frac{2}{1155}v_{dddd\cdot dddd}^6
+\frac{79}{15015}v_{dddd\cdot dddd}^8
-\frac{1}{63\sqrt{11}}v_{dddd_2\cdot dddd_4}^2
+\frac{1}{7\sqrt{1430}}v_{dddd_2\cdot dddd_4}^4,
\nonumber\\&\textstyle
a^{(4)}_{12}=
\frac{1}{462}v_{dddd_4\cdot dddd_4}^2
-\frac{12}{5005}v_{dddd_4\cdot dddd_4}^4
-\frac{1}{210}v_{dddd\cdot dddd}^5
+\frac{1}{165}v_{dddd\cdot dddd}^6
-\frac{16}{15015}v_{dddd\cdot dddd}^8
\nonumber\\&\textstyle\phantom{a^{(4)}_{12}}
+\frac{1}{21\sqrt{11}}v_{dddd_2\cdot dddd_4}^2
-\frac{3}{7\sqrt{1430}}v_{dddd_2\cdot dddd_4}^4.
\label{e_sdclimit4}
\end{align}
States with up to three $d$ bosons are uniquely characterised by their angular momentum $L$
and there is no need to specify the seniority $\upsilon$.
This is no longer the case for $n=4$ where for $L=2$ and 4 the seniority can be $\upsilon=2$ or 4.
This is indicated in Eq.~(\ref{e_sdclimit4}) with an index, that is,
$dddd_\upsilon$ stands for a normalised state of four $d$ bosons with seniority $\upsilon$.


\begin{thebibliography}{[11]}
\bibitem{Iachello95}
F.~Iachello, R.D.~Levine,
{\it Algebraic Theory of Molecules}
(Oxford University Press, Oxford, 1995).

\bibitem{Iachello87}
F.~Iachello, A.~Arima,
{\it The Interacting Boson Model}
(Cambridge University Press, Cambridge, 1987).

\bibitem{Arima77}
A.~Arima, T.~Ohtsuka, F.~Iachello, I.~Talmi,
{\it Collective nuclear states as symmetric couplings of proton and neutron excitations},
Phys.\ Lett.\ B  {\bf66} (1977) 205.

\bibitem{Elliott80}
J.P.~Elliott, A.P.~White,
{\it An isospin invariant form of the interacting boson model},
Phys.\ Lett.\ B {\bf97} (1980) 169.

\bibitem{Shalit63}
A.~de-Shalit, I.~Talmi,
{\it Nuclear Shell Theory}
(Academic Press, New York, 1963).

\bibitem{Talmi93}
I.~Talmi,
{\it Simple Models of Complex Nuclei.
The Shell Model and Interacting Boson Model}
(Harwood, Chur, 1993).

\bibitem{Skouras86}
L.D.~Skouras, S.~Kossionides, 
{\it Generalized fractional parentage coefficients for shell-model calculations},
Comp.\ Phys.\ Comm.\ {\bf39} (1986) 197.

\bibitem{Racah43}
G.~Racah,
{\it Theory of complex spectra},
Phys.\ Rev.\  {\bf63} (1943) 367.

\bibitem{Zerguine11}
S.~Zerguine, P.~Van~Isacker,
{\it Spin-aligned neutron-proton pairs in $N=Z$ nuclei},
Phys.\ Rev.\ C {\bf83} (2011) 064314.

\bibitem{Wybourne70}
B.G.~Wybourne,
{\it Symmetry Principles and Atomic Spectroscopy}
(Wiley-Interscience, New York, 1970).

\bibitem{Weyl39}
H.~Weyl,
{\it The Classical Groups}
(Princeton University Press, Princeton, 1939).

\bibitem{Gheorghe04}
A.~Gheorghe, A.A.~Raduta,
{\it New results for the missing quantum numbers labelling the quadrupole and octupole boson basis}
J.\ Phys.\ A: Math.\ Gen.\ {\bf37} (2004) 10951.

\bibitem{Devi92}
Y.D.~Devi, V.K.B.~Kota,
Pramana J.\ Phys.\  {\bf39} (1992) 413.

\bibitem{Arima78}
A.~Arima, F.~Iachello,
{\it Interacting boson model of collective nuclear states II. The rotational limit},
Ann.\ Phys.\ (NY) {\bf111} (1978) 201.

\bibitem{Scholten79}
O.~Scholten,
KVI Internal report no. 63, 1979.

\bibitem{Scholten80}
O.~Scholten,
{\it The Interacting Boson Approximation Model and Applications},
Ph.D.\ Thesis, University of Groningen, The Netherlands, 1980.

\bibitem{Isacker_ibm.f}
P.~Van~Isacker,
Fortran code ibm.f.

\bibitem{Garcia-Ramos14}
J.E.~Garc\'\i a-Ramos, K.~Heyde,
{\it Nuclear shape coexistence:
A study of the even--even Hg isotopes using the interacting boson model with configuration mixing},
Phys.\ Rev.\ C {\bf89} (2014) 014306.

\bibitem{Heinze08}
S.~Heinze,
{\it Eine Methode zur L\"osung beliebiger bosonischer und fermionischer Vielteilchensysteme},
Ph.D.\ Thesis,  University of Cologne, Germany, 2008.

\bibitem{Brown01}
B.A.~Brown,
{\it The nuclear shell model towards the drip lines},
Prog.\ Part.\ Nucl.\ Phys.\ {\bf47} (2001) 517.

\bibitem{Brown14}
B.A.~Brown, W.D.M.~Rae,
{\it The shell-model code NuShellX@MSU},
Nucl.\ Data Sheets {\bf120} (2014) 115.

\bibitem{Caurier05}
E.~Caurier, G.~Mart\'\i nez-Pinedo, F.~Nowacki, A.~Poves, A.P.~Zuker,
{\it The shell model as a unified view of nuclear structure},
Rev.\ Mod.\ Phys.\ {\bf77} (2005) 427.

\bibitem{Shimizu12}
N.~Shimizu, T.~Abe, Y.~Tsunoda, Y.~Utsuno, T.~Yoshida, T.~Mizusaki, M.~Honma, T.~Otsuka,
{\it New-generation Monte Carlo shell model for the K computer era},
Prog.\ Theor.\ Exp.\ Phys.\ {\bf 2012}, 1, 01A205.

\bibitem{Shimizu19}
Shimizu, N.; Mizusaki, T.; Utsuno, Y.; Tsunoda, Y.
Thick-restart block Lanczos method for large-scale shell-model calculations.
{\it Comp.\ Phys.\ Comm.\ } {\bf244} (2019)  372.

\bibitem{Johnson13}
C.W.~Johnson, W.E.~Ormand, P.G.~Krastev,
{\it Factorization in large-scale many-body calculations},
Comp.\ Phys.\ Comm.\ {\bf184} (2013) 2761.

\bibitem{Johnson18}
C.W.~Johnson, W.E.~Ormand, K.S.~McElvain, H.~Shan,
{\it BIGSTICK: A flexible configuration-interaction shell-model code},
arXiv:1801.08432.

\bibitem{Isacker_ibm.m}
P.~Van~Isacker,
Mathematica code {\tt ibm.m}.

\bibitem{Frank94}
A.~Frank, P.~Van~Isacker,
{\it Algebraic Methods in Molecular and Nuclear Structure Physics}
(Wiley-Interscience, New York, 1994).

\bibitem{Ginocchio80}
J.N.~Ginocchio, M.W.~Kirson,
{\it Relationship between the Bohr collective Hamiltonian and the interacting-boson model},
Phys.\ Rev.\ Lett.\ {\bf44} (1980) 1744.

\bibitem{Dieperink80}
A.E.L.~Dieperink, O.~Scholten, F.~Iachello,
{\it Classical limit of the interacting-boson model},
Phys.\ Rev.\ Lett.\ {\bf44} (1980) 1747.

\bibitem{Bohr80}
A.~Bohr, B.R.~Mottelson,
{\it Features of nuclear deformations produced by the alignment of individual particles or pairs},
Phys.\ Scripta {\bf22} (1980) 468.

\bibitem{Isacker81}
P.~Van~Isacker, J.-Q.~Chen,
{\it Classical limit of the interacting boson Hamiltonian},
Phys.\ Rev.\ C {\bf24} (1981) 684.

\bibitem{Fortunato11}
L.~Fortunato, C.E.~Alonso, J.M.~Arias, J.E.~Garc\'\i a-Ramos, A.~Vitturi,
{\it Phase diagram for a cubic-$Q$ interacting boson model Hamiltonian: Signs of triaxiality},
Phys.\ Rev.\ C {\bf84} (2011) 014326.

\bibitem{Poves20}
A.~Poves, F.~Nowacki, Y.~Alhassid,
{\it Limits on assigning a shape to a nucleus},
Phys.\ Rev.\ C {\bf101} (2020) 054307.

\bibitem{Warner83}
D.D.~Warner, R.F.~Casten,
{\it Predictions of the interacting boson approximation in a consistent $Q$ framework},
Phys.\ Rev.\ C {\bf28} (1983) 1798.

\bibitem{Sorgunlu08}
B.~Sorgunlu, P.~Van~Isacker,
{\it Triaxiality in the interacting boson model},
Nucl.\ Phys.\ A {\bf808} (2008) 27.

\end{thebibliography}
\end{document}